\documentclass[aps,prl,twocolumn,showpacs,groupedaddress,superscriptaddress,footinbib,longbibliography]{revtex4-2}

\usepackage{pdfpages} 
\usepackage{pgffor} 

\makeatletter
\AtBeginDocument{\let\LS@rot\@undefined}
\makeatother

\def\supplementfilename{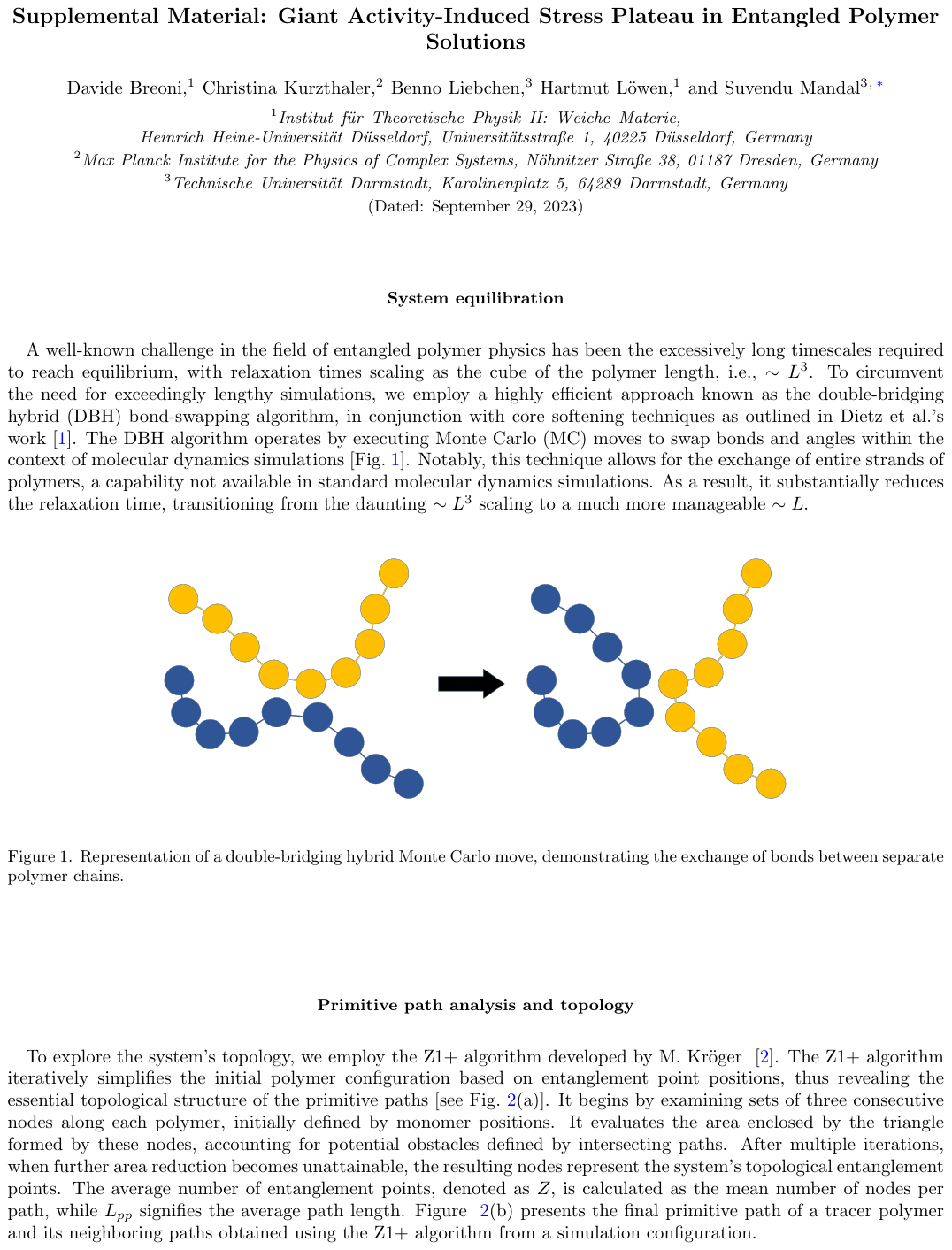}

\pdfximage{\supplementfilename}
\def\numbersupplementpages{\the\pdflastximagepages}

\newif\ifarXiv
\arXivtrue 
\usepackage{textcomp}

\usepackage{mathrsfs}
\usepackage{natbib,hyperref}
\setcitestyle{square,sort&compress,comma,numbers}
\hypersetup{colorlinks=true, citecolor=blue, urlcolor=blue, linkcolor=blue}
\usepackage[english]{babel}
\usepackage[utf8]{inputenc}
\usepackage{graphicx}
\usepackage[caption=false]{subfig}
\usepackage{amsmath}
\usepackage{amsfonts}
\usepackage{amssymb}
\usepackage{color,soul}
\setulcolor{red}
\usepackage{easyReview}
\usepackage{xcolor}

\usepackage[normalem]{ulem}

\usepackage{newunicodechar}

\begin{document}

\title{Giant Activity-Induced Stress Plateau in Entangled Polymer Solutions}
\author{Davide Breoni}
\affiliation{Institut f\"ur Theoretische Physik II: Weiche Materie, Heinrich
Heine-Universit\"at D\"usseldorf, Universit\"atsstra{\ss}e 1, 
40225 D\"usseldorf, Germany}
\author{Christina Kurzthaler}
\affiliation{Max Planck Institute for the Physics of Complex Systems, N\"ohnitzer Stra{\ss}e 38,
01187 Dresden, Germany}
\author{Benno Liebchen}
\affiliation{Technische Universit\"at Darmstadt, Karolinenplatz 5, 64289 Darmstadt, Germany}
\author{Hartmut L\"owen}
\affiliation{Institut f\"ur Theoretische Physik II: Weiche Materie, Heinrich
Heine-Universit\"at D\"usseldorf, Universit\"atsstra{\ss}e 1, 
40225 D\"usseldorf, Germany}
\author{Suvendu Mandal}
\email{suvendu.mandal@pkm.tu-darmstadt.de}
\affiliation{Technische Universit\"at Darmstadt, Karolinenplatz 5, 64289 Darmstadt, Germany}

\begin{abstract}
We study the viscoelastic properties of highly entangled, flexible, self-propelled polymers using Brownian dynamics simulations. Our results show that the active motion of the polymer increases the height of the stress plateau by orders of magnitude due to the emergence of grip forces at entanglement points. Identifying the activity-induced energy of a single polymer and the ratio of polymer length to self-propulsion velocity as relevant energy and time scales, we find the stress autocorrelation functions collapse across P\'eclet numbers. We predict that the long-time viscosity scales with polymer length squared $\sim L^2$, in contrast to equilibrium counterparts $\sim L^3$. These insights offer prospects for designing new materials with activity-responsive mechanical properties.

\end{abstract} 
\maketitle

Entangled polymer solutions represent fundamental building blocks of many biological materials, where they serve functions as diverse as cell mitosis~\cite{howard_mechanics_2005,bausch_bottom-up_2006,schaller_polar_2010,sanchez_spontaneous_2012,alberts_molecular_2017} and transcription of genetic material~\cite{Levi:2005,Zidovska:2013,Saintillan:2018}. Furthermore, they are important collective life forms, which provide individuals resistance to environmental stresses~\cite{deblais_rheology_2020,deblais_phase_2020,ozkan-aydin_collective_2021,Patil:2023science}, and lay the foundation for numerous technological applications~\cite{Yardimci2023bonded,Patil:2023science,Deblais2023worm}. The rheological properties of these complex materials are governed by the elasticity and structure of their conformations, such as their long, slender linear~\cite{Wingstrand:2015}, twisted~\cite{Smrek:2021}, or loop conformations~\cite{Kapnistos:2008}, their strong entanglement, and their specific microscopic interactions, which makes them a fascinating many-body problem in physics. 

The viscoelastic properties of these strongly-interacting systems at thermodynamic equilibrium have been thoroughly studied in the realm of polymer physics. A major breakthrough has been the theoretical prediction of rheological properties of entangled linear polymer melts in terms of their stress  autocorrelation function, which exhibits a prominent plateau at intermediate times, characterizing the elastic response, and relaxes exponentially at long times~\cite{doi_theory_1986, rubinstein_polymer_2003,deGennes:1971}. The relation between phenomenological parameters of the underlying tube model and microscopic system properties to ultimately predict the stress plateau has been established by analyzing the polymers' primitive paths~\cite{everaers_rheology_2004,Hoy:2020PRL}, which correspond to the axes of entangled polymer tubes~\cite{everaers_rheology_2004,Hoy:2020PRL}. While it has been shown that the stress plateau of linear polymer solutions remains unaffected by external driving~\cite{hou_stress_2010}, tuning the topological properties of the polymers can lead to a qualitative change of the stress relaxation dynamics~\cite{Kapnistos:2008}. The latter display a power-law behavior for loop polymer melts and recover a stress plateau only upon adding linear polymer chains to the solution~\cite{Kapnistos:2008}. 

Recent work has demonstrated that microscopic interactions among the entangled constituents can be governed by active components, such as molecular motors in solution~\cite{schaller_polar_2010,sanchez_spontaneous_2012,Levi:2005,Zidovska:2013,Saintillan:2018} or the intrinsic motility of the individuals~\cite{deblais_rheology_2020,deblais_phase_2020,ozkan-aydin_collective_2021,Patil:2023science}, which drive these systems far from equilibrium and generate dynamical and structural behaviors distinct from their passive counterparts. Understanding the interplay of entanglement and activity is not only fundamental to living systems but also crucial for designing and processing new soft materials with tailored properties. In particular, incorporating active components in addition to tuning the entanglement has the potential to improve the mechanical properties of materials. Yet, theoretical studies in this direction are limited and no universal behaviors or scaling predictions have been established to guide experimental progress.

Here, we use Brownian dynamics simulations to characterize the viscoelastic properties of  highly-entangled, flexible, self-propelled polymers in terms of the time-dependent stress autocorrelation function and viscosity. Our results reveal a remarkable amplification of the stress plateau, a phenomenon intricately linked to the interplay of active motion and topological uncrossability of polymers, leading to the emergence of grip forces. In particular, neighboring polymers form hairpin structures that exert forces, pulling the entangled test polymer in the direction of their self-propulsion, effectively preventing its sliding at the entanglement points. It is noteworthy that the magnitude of these grip forces depends on the self-propulsion velocity. Subsequently, we show that the stress autocorrelation functions for a broad range of polymer lengths and P{\'e}clet numbers can be collapsed onto a single master curve by identifying the relevant energy and time scales. Finally, we predict that the long-time viscosity scales with the square of the polymer length $\sim L^2$, which becomes exact for high P{\'e}clet numbers in the highly-entangled regime.

\begin{figure*}[tp]
    \centerline{\includegraphics[width=1.25\textwidth]{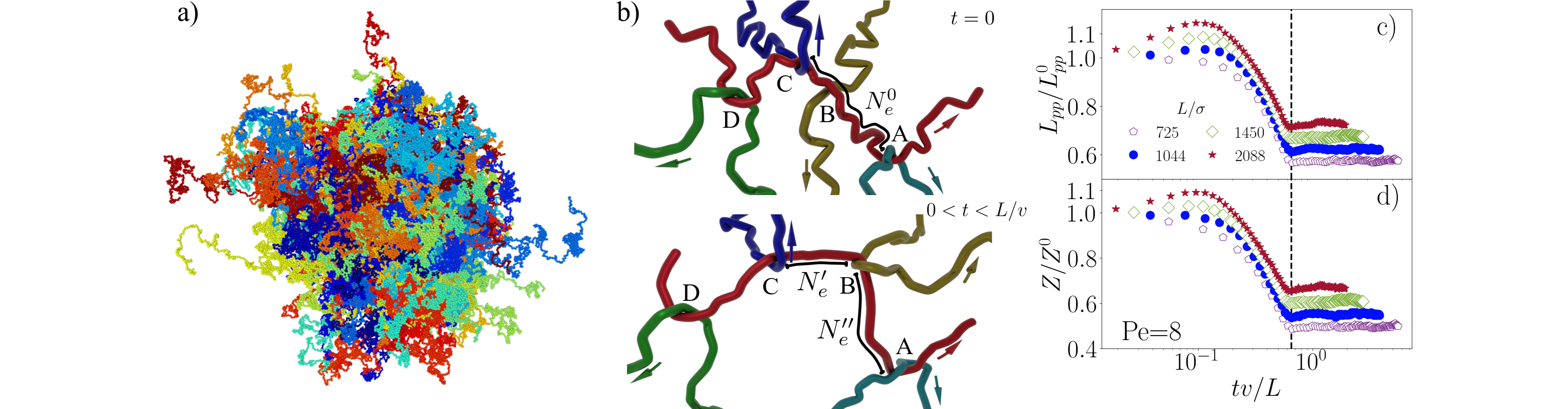}}
	\caption{(a) Simulation snapshot of entangled, flexible polymers (each polymer has its own color). (b) 3D illustration depicting the primitive path of a test polymer (red line) confined within an effective tube formed by surrounding self-propelled polymers at various times $t$. In the equilibrium state ($t=0$), a combination of strong entanglement points (A, C, and D) and weak entanglement point (B) coexists, with strong entanglements distinguished by the presence of hairpin structures. Due to activity, prior to reaching the steady state ($t \gtrsim L/v$), the number of strong entanglement points increases (as shown by the yellow polymer wrapping around the red polymer at point $B$), resulting in the elongation of the primitive path. The direction of self-propulsion is indicated by colored arrows, while the distance between successive entanglement points defines the entanglement length $N_e$. (c)~Contour length of the primitive path $L_{pp}$, normalized by the equilibrium primitive path $L_{pp}^0$ as a function of time for different polymer lengths $L$ and fixed P{\'e}clet number ${\rm Pe}=8$. Time is rescaled by the ratio of polymer length to self-propulsion velocity $L/v$. (d) Number of entanglement points $Z$, normalized by the number of entanglement points $Z^0$ for ${\rm Pe}=0$, as a function time.}
    \label{fig:Fig1}
\end{figure*}

\paragraph{Model--}
We perform 3D Brownian dynamics simulations of highly-entangled polymer solutions of $N$ self-propelled, flexible polymer chains using the bead-spring model~\cite{Kremer:1990model}. Each chain consists of $N_p$ monomers with diameter $\sigma$ and has a length of $L=N_p\sigma$. The connectivity and repulsion of the beads are modeled using the finitely extensible nonlinear elastic potential (FENE)~\cite{Kremer:1990model} and the Weeks-Chandler-Andersen potential (WCA)~\cite{Weeks:1971wca} with energies $\epsilon_{\rm FENE}$ and $\epsilon_{\rm WCA}$, respectively. Angular interactions along chain backbones are captured using a bending potential for each monomer $U_{\text{ang}, i}=\kappa \sum_{j=i-1,i,i+1} (1 -\textbf{t}_j \cdot \textbf{t}_{j+1})$, where $\textbf{t}_j =(\textbf{r}_{j+1} - \textbf{r}_{j})/(|\textbf{r}_{j+1} - \textbf{r}_{j}|)$ represents the tangent vector between consecutive monomers having positions $\mathbf{r}_{j}$ and $\kappa$ corresponds to the bending energy. The polymers are subject to Brownian motion modeled by stochastic forces $\textbf{F}_{r,i}$, where $\langle F_{r,i}^{\alpha}(t)F_{r,j}^{\beta}(t')\rangle=2k_BT\zeta \delta_{ij}\delta_{\alpha\beta} \delta(t'-t)$ with friction coefficient $\zeta$ and thermal energy $k_BT$. Their self-propulsion is modeled by an active force $\textbf{F}_{p,i}$ acting tangentially to the polymer contour~\cite{kurzthaler_geometric_2021,Bianco:2018,Prathyusha:2018}, so that (without interactions) each bead moves at a velocity of $v=|\mathbf{F}_{p,i}|/\zeta$ ($|\mathbf{F}_{p,i}|$ being constant across all monomers). Thus, the equation of motion for each monomer read
\begin{equation}
\label{MotEq}
\zeta\frac{\text{d}\textbf{r}_i}{\text{d}t}= -\nabla_i U+\textbf{F}_{p,i}+\textbf{F}_{r,i}.
\end{equation}

Dimensionless parameters, derived from length and time units ($\sigma$ and $\tau_0=\sigma^2/D_0$, with $D_0=k_BT/\zeta$ as the short-time diffusion coefficient of a monomer), include the P\'eclet number ($\text{Pe} = v \sigma/D_0$) for assessing the significance of active motion relative to diffusion, along with coupling parameters ($\epsilon_{\rm WCA}/k_BT$, $\epsilon_{\rm FENE}/k_BT$, and $\kappa/k_B T$). Additionally, we define the dimensionless density $\rho^{\star}=N_{\rm tot} \sigma^3/V$, where $V$ denotes the volume of the simulation box. We keep fixed values of $\rho^{\star}=0.85$, $\epsilon_{\rm WCA}/k_BT=1.0$, $\epsilon_{\rm FENE}/k_BT=30$, and $\kappa/k_B T=1.0$, while systematically varying the polymer length ($L/\sigma=100,...2088$), resulting in a dimensionless entanglement length $N_e \cong 41$~\cite{kroger_z1_2023}. Equations of motion are solved numerically using a modified version of LAMMPS with a time step of $\delta t = 10^{-4}\tau_0$. Equilibration is achieved through a bond-swapping algorithm with core softening [see SI~\cite{suppl}], and all time measurements are referenced from this equilibration point. Notably, both active and passive highly-entangled polymer systems exhibit an ideal chain scaling relation for the end-to-end distance $R_{ee}\propto L^{1/2}$, in contrast to dilute active polymer solutions~\cite{Bianco:2018}, indicating that activity does not affect this scaling [see SI~\cite{suppl}].

\begin{figure*}[tp]
    \centerline{\includegraphics[width=1.22\textwidth]{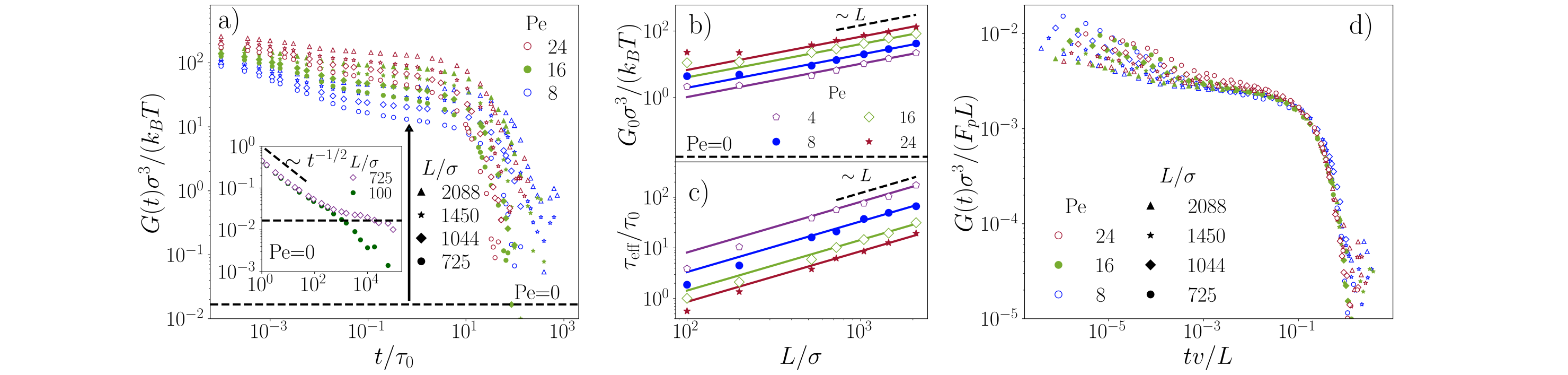}}
	\caption{(a) Stress autocorrelation function $G(t)$ for different P{\'e}clet numbers and polymer lengths as a function of time. Inset: The stress autocorrelation function in equilibrium for $L=725 \sigma$ validates the well-established prediction $G_0=4\rho k_BT/(5N_e^0)$ (dashed line), where $N_e^0$ is the entanglement length between two subsequent entanglement points. Axes are labeled as in the main figure. Black arrow indicates the giant activity-induced stress plateau compared to the equilibrium state. (b)~Stress plateau $G_0$ and (c)~disengagement time $\tau_{\rm eff}$ as a function of polymer length $L$ extracted from our simulations for a wide range of P{\'e}clet numbers. (d)~Rescaled stress autocorrelation function $G(t)\sigma^3/F_p L$ as a function of the rescaled time $tv/L$.}
	\label{fig:stress}
\end{figure*}

\paragraph{Activity-enhanced stress plateau--} The viscoelastic properties of polymer solutions are encoded in the stress autocorrelation function 
\begin{align}
G(t)=\frac{V}{3k_BT}\sum_{\alpha\neq\beta}\left\langle \sigma_{\alpha \beta}(t)\sigma_{\alpha \beta}(0)\right\rangle,
\end{align}

where the sum runs over all off-diagonal components of the stress tensor $\sigma_{\alpha\beta}$ and $\langle .. \rangle$ denotes an ensemble average. In equilibrium systems, for the case of short, unentangled linear polymer solutions, this yields the power-law dynamics described by the Rouse model, $G(t)\sim t^{-1/2}$~\cite{rubinstein_polymer_2003}. In contrast, highly-entangled polymers are forced to move along the direction of their contour, while their motion perpendicular to it is restricted to a tube-like region formed by the surrounding polymers [Fig.\ref{fig:Fig1}(a)]. Consequently, the stress autocorrelation function exhibits a plateau $G_0$ at intermediate times and an exponential decay $G_0\text{e}^{-t/\tau_{\rm eff}}$ at long times. The stress plateau $G_0$, a hallmark of entangled polymer chains, quantifies the elasticity of the system, while the disengagement time $\tau_{\rm eff}\sim L^3$ corresponds to the characteristic time the polymer requires to move its own length~$L$ along the tube.

To investigate the effect of activity, we compute the stress autocorrelation function $G(t)$ for self-propelled polymers of different lengths, $L=100,...2088\sigma$, and P\'eclet numbers, $\rm Pe$~$=1,...24$, see Fig.~\ref{fig:stress}(a). At very short times ($t\lesssim 10^{-3}\tau_0$), the active polymer solution is slightly harder ($G(t)$ increases by a factor of $4$ compared to the passive counterpart [see SI~\cite{suppl}]), which can be attributed to the increased fluctuations exhibited by the self-propelled polymers within their tubes.

At intermediate times, $t\sim \tau_0$, the difference between the stress autocorrelation function $G(t)$ of passive and active systems becomes significantly larger, which becomes apparent in an increase of the plateau height $G_0$ by three orders in magnitude [see Fig.~\ref{fig:stress}(a)]. This amplification arises from grip forces exerted on the red test polymer by neighboring polymers [Fig.\ref{fig:Fig1}(b)]. First, these neighboring polymers form hairpin structures around the test polymer, stretching its primitive path, thereby slowing down the relaxation of $G(t)$ as the test polymer traverses within an elongated tube. This effect becomes pronounced when we keep the P\'eclet number constant while increasing the polymer length [Fig.\ref{fig:stress}(b)]. Second and more strikingly, these grip forces also act as barriers, effectively preventing the test polymer from sliding at the entanglement points. This results in a substantial increase in the plateau height as $\rm Pe$ increases at a fixed polymer length [Fig.\ref{fig:stress}(a) and SI~\cite{suppl}]. Hence, both mechanisms contribute to a giant enhancement of the elastic stress plateau height, a phenomenon exclusive to self-propelled entangled systems. When the test polymer disengages from its tube, the grip forces imposed by the surrounding polymers diminish, leading to a  relaxation of the stress autocorrelation function from the plateau at long times [see Fig.~\ref{fig:stress}(a)].

This physical picture can be corroborated by measuring the average contour length of the primitive path, denoted as $L_{pp}$, and the average number $Z$ of entanglement points. To elucidate topological entanglement dynamics, we employed the Z1+ topological analysis algorithm~\cite{kroger_z1_2023,kroger_shortest_2005,shanbhag_primitive_2007,hoy_topological_2009,karayiannis_combined_2009}, which systematically undergoes a sequence of geometric minimizations. The primitive path is rigorously defined as the shortest path between the two ends of a polymer chain while preserving its topological uncrossability. At intermediate times $tv/L \sim 0.1$, our simulations show that upon increasing the polymer length at a fixed $\text{Pe}=8$, $L_{pp}$ and $Z$ increase by $10\%$ compared to the passive counterpart [see Fig.~\ref{fig:Fig1}(c-d)]. This observation suggests that the active system becomes more entangled, with the number of entanglement points rising from $Z=105$ to $115$ for $L/\sigma=2088$. Moreover, we evaluated the entanglement length $N_e$ using the relation $N_e=(N_p -1)\langle R_{ee}^2 \rangle /\langle L_{pp}^2 \rangle$~\cite{hoy_topological_2009,kroger_z1_2023}. In contrast to $L_{pp}$, the end-to-end distance $R_{ee}$ exhibits a gradual decrease until it eventually saturates at long times ($tv/L \gg 1$) at a fixed $\rm Pe=8$ [see SI~\cite{suppl}]. Consequently, at intermediate times ($tv/L \sim 0.1$), we observe a reduction of approximately $30\%$ in $N_e$ relative to the passive counterpart [see SI~\cite{suppl}].

It is tempting to validate the giant increase in the stress plateau $G_0$ via the well-established relation for equilibrium systems $G_0=4\rho k_BT/(5N_e)$~\cite{hsu_static_2016}. However, our observations reveal a $30\%$ decrease in $N_e$ with increasing polymer length $L$ at a fixed P{\'e}clet number (${\rm Pe}=8$), while the stress plateau $G_0$ increases by orders of magnitude. By employing a dimensional argument, we show that the enhanced stress plateau can rather be related to the active energy of a single polymer $F_p L$, where $F_p$ denotes the magnitude of the active force. For large ${\rm Pe}\gg 1$, this energy dominates over thermal energy and thus represents the relevant energy scale of our system, leading to our prediction $G_0\sim F_pL/\sigma^3$. To quantify this phenomenon, we show the plateau height $G_0$ as a function of the polymer length $L$ for a range of P{\'e}clet numbers in Fig.~\ref{fig:stress}(b). It turns out that $G_0$ indeed increases linearly as a function of the polymer length in the highly entangled regime ($L \gtrsim 500\sigma$). This occurs since the polymers are forced to move within elongated tubes as well as the system gets highly entangled (the number of entanglement points $Z$ increases compared to the passive counterpart).  However, for unentangled chains with $L\lesssim 100 \sigma$, the stress plateau vanishes and we recover an algebraic decay $\sim t^{-1/2}$, in agreement with  the Rouse model, which validates the idea that the stress plateau is a unique feature of highly entangled polymer solutions (see SI~\cite{suppl}).

At long times $t\gg \tau_0$, the stress autocorrelation function follows the expected exponential decay $G(t)\sim G_0\exp(-t/\tau_{\rm eff})$, where $\tau_{\rm eff}$ represents the disengagement time of our active system [see Fig.~\ref{fig:stress}(a)]. At these times, the transverse motion becomes nearly frozen, allowing the polymer to self-propel and diffuse freely along the tube at timescales of $L/v$ and $\sim L^3$, respectively. The disengagement time $\tau_{\rm eff}$ is determined by the faster of these two mechanisms and we use the interpolation formula given below as an estimate:
\begin{align}
\tau_\text{eff}^{-1}=D_0\sigma/L^3 + v/L.
\end{align}
Remarkably, our computer simulations show that active entangled polymers relax much faster than their passive counterparts, resulting in a disengagement time that scales as $\tau_\text{eff} \sim L$  [see Fig.\ref{fig:stress}(c)]. This is in contrast to the passive case, where the disengagement time scales as $\sim L^3$ for larger polymer lengths, as observed in experiments~\cite{Keshavarz2016:nanoscale}. 

By combining the relevant time $\tau_{\rm eff}\sim L/v$ and energy scales $G_0 \sim F_pL/\sigma^3$, the data collapse onto a single curve at intermediate and long times, as depicted in Fig.\ref{fig:stress}(d). The data collapse is excellent over nearly three decades in time, confirming our predictions. 

\begin{figure*}[tp]
 \begin{center}
\centerline{\includegraphics[width=1.24\textwidth]{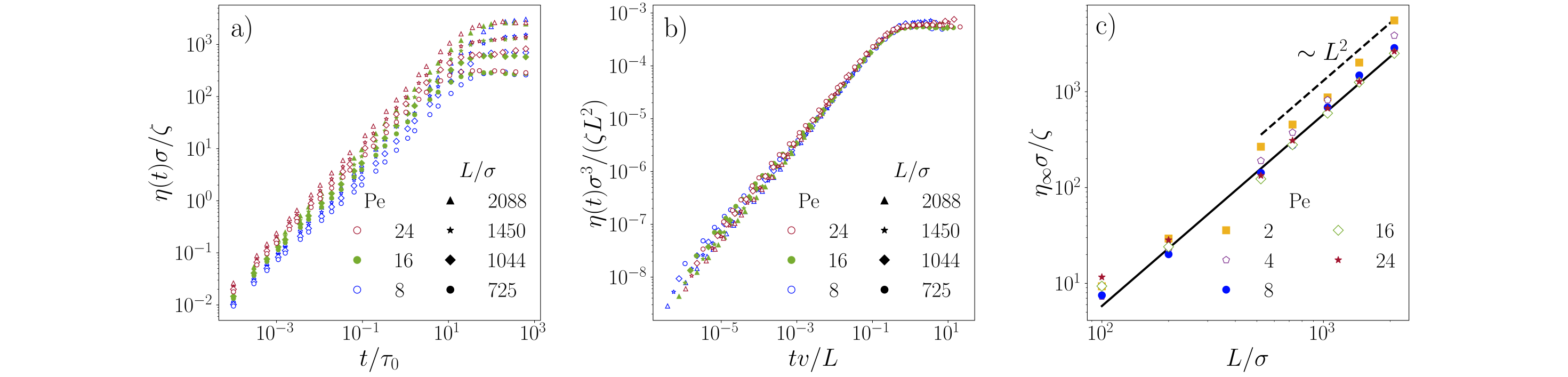}}
 \caption{(a) Time-dependent viscosity for a wide range of P{\'e}clet numbers and polymer lengths. (b) A data collapse is obtained by rescaling the viscosity by $\eta(t) \sigma^3/\zeta L^2$ and the time scale by $tv/L$. (c) Long-time viscosity $\eta_{\infty}$ as a function of polymer length $L$ extracted from simulations for a wide range of P{\'e}clet numbers. The black line indicates the scaling of  $\eta_\infty \sim L^2$.}
 \label{fig:visco}
 \end{center}
\end{figure*}

\paragraph{Time-dependent viscosity--} 
Following our previous predictions ($G_0\sim LF_p/\sigma^3$ and $\tau_\text{eff}\sim L/v$), the time-dependent and stationary viscosity are expected to scale as $\eta \sim G_0\tau_{\rm eff} \sim L^2$. Only recently, it has been claimed that in the hydrodynamic limit (i.e., at long times and at large length scales) the Green-Kubo relation is valid even for suspensions of active dumbells~\cite{Han_fluctuating:2021}. This work inspired us to use the Green-Kubo relation, offering access to the time-dependent viscosity of our entangled system via 
\begin{align}
\eta(t)&=\int_0^t G(t')dt',
\end{align}
which is shown in Fig.~\ref{fig:visco}(a) over 6 decades in time. Our study suggests that at short times $t\lesssim\tau_0$, activity and entanglement play a minor role, but at intermediate times the data become significantly different. Following our previous predictions ($G_0\sim LF_p/\sigma^3$ and $\tau_\text{eff}\sim L/v$), the time-dependent and stationary viscosity are expected to scale as $\eta \sim G_0\tau_{\rm eff} \sim L^2$. Rescaling the data accordingly, we find a collapse onto a single master curve over 4 orders of magnitude in time [Fig.~\ref{fig:visco}(b)].

Finally, as our data saturate at long times we can estimate the stationary viscosity of the system via $\eta_\infty \equiv \lim_{t\rightarrow \infty}\eta(t)$. First, we find that the stationary viscosity remains independent of $\rm Pe$ when the polymer length is fixed [see Fig.~\ref{fig:visco}(a)]. Second and more strikingly, the predicted scaling $\eta _\infty \sim L^2$ is confirmed by an asymptotic data collapse in the regime of high entanglement ($L\gtrsim 500 \sigma$) and high P\'eclet numbers ($\text{Pe} \gtrsim 8$) [Fig.~\ref{fig:visco}(c)]. Hence, highly-engtangled active solutions follow a generic scaling of $~L^2$, which is distinct from the characteristic $~L^3$ scaling that broadly applies to equilibrium systems. Deviations become apparent for shorter polymer lengths ($L \le 250$), where the solution becomes less entangled. This can be attributed to the fact that as we increase the P{\'e}clet number, the tube diameter ($\sim \sqrt{N_e}\sigma$)~\cite{rubinstein_polymer_2003} also becomes larger. Therefore, it requires even longer polymers to observe a highly-entangled state. 

\paragraph*{Conclusions--}

Our study reveals a profound impact of activity on entangled polymer solutions, notably enhancing the stress plateau height, and predicts a scaling law for the stationary viscosity $\eta_{\infty} \sim L^2$, which goes beyond common knowledge and contrasts with the characteristic $\eta_\infty \sim L^3$ law for equilibrium systems.

Our findings open up new avenues for quantifying the viscoelastic properties of various experimental systems. While on the macroscale the dynamics of highly entangled \textit{T. Tubifex} worms~\cite{Deblais2023worm} could be studied, on the microscale activated nanotubes~\cite{Bacanu_activeNanotubes:2023}, synthetic polymer chains~\cite{Yang_reconfigurable:2020}, or chromatin~\cite{Saintillan:2018} represent potential realizations for entangled systems with unique properties. Our framework can provide insights for systems under deformation/shear, which may allow measuring the material properties of these systems in the presence of another time scale (inverse shear rate).

While significant progress has been made in understanding the viscoelastic properties of passive entangled systems under deformation~\cite{vanmacromolecular:2022,Hsuclustering:2019,Wang:2007new,Wang:2007new,Cates1993:rheology}, it is essential to highlight two key distinctions: First, active entangled systems exhibit a remarkable increase in the stress plateau height, whereas deformed passive entangled polymers typically experience a reduction in the stress plateau height~\cite{vanmacromolecular:2022,Hsuclustering:2019}. Second, active entangled systems remain force-free and do not develop a finite stress, contrasting with the behavior of deformed passive entangled systems~\cite{OConnorstress:2019}.

Moreover, our study focuses on self-propelled flexible polymers, yet many polymers found in nature are semiflexible~\cite{Banerjee:2020actin,winkler_physics_2020,Eisenstecken:2017,Eisenstecken:2022path,Harnau:1996dynamic,Kroy:1996,Hallatschek:2005,Lang:2018disentangling,mandal_crowding-enhanced_2020}. Therefore, a future challenge is to include the finite bending rigidity of polymers in our analysis and explore how the stress plateau and disengagement time vary with swimming speed. This would deepen our understanding of the behavior of biological filaments and contribute to the development of advanced materials with tailored viscoelastic properties, such as synthetic cells~\cite{Blain:2014,Ganzinger:2019}.

\begin{acknowledgments}
S.M. gratefully acknowledges Robert S. Hoy and Joseph D. Dietz for valuable discussions regarding the equilibration of entangled polymer solutions, as well as Martin Kr\"oger for sharing the topological analysis code before publishing it. The work of D.B. was supported within the EU MSCA-ITN ActiveMatter (proposal No.\ 812780). H.L. acknowledges funds from the German Research Foundation (DFG) project LO 418/29-1.
\end{acknowledgments}


%


\begin{thebibliography}{55}%
\makeatletter
\providecommand \@ifxundefined [1]{%
 \@ifx{#1\undefined}
}%
\providecommand \@ifnum [1]{%
 \ifnum #1\expandafter \@firstoftwo
 \else \expandafter \@secondoftwo
 \fi
}%
\providecommand \@ifx [1]{%
 \ifx #1\expandafter \@firstoftwo
 \else \expandafter \@secondoftwo
 \fi
}%
\providecommand \natexlab [1]{#1}%
\providecommand \enquote  [1]{``#1''}%
\providecommand \bibnamefont  [1]{#1}%
\providecommand \bibfnamefont [1]{#1}%
\providecommand \citenamefont [1]{#1}%
\providecommand \href@noop [0]{\@secondoftwo}%
\providecommand \href [0]{\begingroup \@sanitize@url \@href}%
\providecommand \@href[1]{\@@startlink{#1}\@@href}%
\providecommand \@@href[1]{\endgroup#1\@@endlink}%
\providecommand \@sanitize@url [0]{\catcode `\\12\catcode `\$12\catcode `\&12\catcode `\#12\catcode `\^12\catcode `\_12\catcode `\%12\relax}%
\providecommand \@@startlink[1]{}%
\providecommand \@@endlink[0]{}%
\providecommand \url  [0]{\begingroup\@sanitize@url \@url }%
\providecommand \@url [1]{\endgroup\@href {#1}{\urlprefix }}%
\providecommand \urlprefix  [0]{URL }%
\providecommand \Eprint [0]{\href }%
\providecommand \doibase [0]{https://doi.org/}%
\providecommand \selectlanguage [0]{\@gobble}%
\providecommand \bibinfo  [0]{\@secondoftwo}%
\providecommand \bibfield  [0]{\@secondoftwo}%
\providecommand \translation [1]{[#1]}%
\providecommand \BibitemOpen [0]{}%
\providecommand \bibitemStop [0]{}%
\providecommand \bibitemNoStop [0]{.\EOS\space}%
\providecommand \EOS [0]{\spacefactor3000\relax}%
\providecommand \BibitemShut  [1]{\csname bibitem#1\endcsname}%
\let\auto@bib@innerbib\@empty
\bibitem [{\citenamefont {Howard}(2005)}]{howard_mechanics_2005}%
  \BibitemOpen
  \bibfield  {author} {\bibinfo {author} {\bibfnamefont {J.}~\bibnamefont {Howard}},\ }\href {https://link.springer.com/book/9780878933334} {\emph {\bibinfo {title} {Mechanics of Motor Proteins and the Cytoskeleton}}}\ (\bibinfo  {publisher} {Springer},\ \bibinfo {year} {2005})\BibitemShut {NoStop}%
\bibitem [{\citenamefont {Bausch}\ and\ \citenamefont {Kroy}(2006)}]{bausch_bottom-up_2006}%
  \BibitemOpen
  \bibfield  {author} {\bibinfo {author} {\bibfnamefont {A.~R.}\ \bibnamefont {Bausch}}\ and\ \bibinfo {author} {\bibfnamefont {K.}~\bibnamefont {Kroy}},\ }\bibfield  {title} {{\selectlanguage {english}\bibinfo {title} {A bottom-up approach to cell mechanics}},\ }\href {https://doi.org/10.1038/nphys260} {\bibfield  {journal} {\bibinfo  {journal} {Nat. Phys.}\ }\textbf {\bibinfo {volume} {2}},\ \bibinfo {pages} {231} (\bibinfo {year} {2006})}\BibitemShut {NoStop}%
\bibitem [{\citenamefont {Schaller}\ \emph {et~al.}(2010)\citenamefont {Schaller}, \citenamefont {Weber}, \citenamefont {Semmrich}, \citenamefont {Frey},\ and\ \citenamefont {Bausch}}]{schaller_polar_2010}%
  \BibitemOpen
  \bibfield  {author} {\bibinfo {author} {\bibfnamefont {V.}~\bibnamefont {Schaller}}, \bibinfo {author} {\bibfnamefont {C.}~\bibnamefont {Weber}}, \bibinfo {author} {\bibfnamefont {C.}~\bibnamefont {Semmrich}}, \bibinfo {author} {\bibfnamefont {E.}~\bibnamefont {Frey}},\ and\ \bibinfo {author} {\bibfnamefont {A.~R.}\ \bibnamefont {Bausch}},\ }\bibfield  {title} {{\selectlanguage {english}\bibinfo {title} {Polar patterns of driven filaments}},\ }\href {https://doi.org/10.1038/nature09312} {\bibfield  {journal} {\bibinfo  {journal} {Nature}\ }\textbf {\bibinfo {volume} {467}},\ \bibinfo {pages} {73} (\bibinfo {year} {2010})}\BibitemShut {NoStop}%
\bibitem [{\citenamefont {Sanchez}\ \emph {et~al.}(2012)\citenamefont {Sanchez}, \citenamefont {Chen}, \citenamefont {DeCamp}, \citenamefont {Heymann},\ and\ \citenamefont {Dogic}}]{sanchez_spontaneous_2012}%
  \BibitemOpen
  \bibfield  {author} {\bibinfo {author} {\bibfnamefont {T.}~\bibnamefont {Sanchez}}, \bibinfo {author} {\bibfnamefont {D.~T.~N.}\ \bibnamefont {Chen}}, \bibinfo {author} {\bibfnamefont {S.~J.}\ \bibnamefont {DeCamp}}, \bibinfo {author} {\bibfnamefont {M.}~\bibnamefont {Heymann}},\ and\ \bibinfo {author} {\bibfnamefont {Z.}~\bibnamefont {Dogic}},\ }\bibfield  {title} {{\selectlanguage {english}\bibinfo {title} {Spontaneous motion in hierarchically assembled active matter}},\ }\href {https://doi.org/10.1038/nature11591} {\bibfield  {journal} {\bibinfo  {journal} {Nature}\ }\textbf {\bibinfo {volume} {491}},\ \bibinfo {pages} {431} (\bibinfo {year} {2012})}\BibitemShut {NoStop}%
\bibitem [{\citenamefont {Alberts}(2017)}]{alberts_molecular_2017}%
  \BibitemOpen
  \bibfield  {author} {\bibinfo {author} {\bibfnamefont {B.}~\bibnamefont {Alberts}},\ }\href@noop {} {\emph {\bibinfo {title} {Molecular {Biology} of the {Cell}}}}\ (\bibinfo  {publisher} {Garland Science},\ \bibinfo {year} {2017})\BibitemShut {NoStop}%
\bibitem [{\citenamefont {Levi}\ \emph {et~al.}(2005)\citenamefont {Levi}, \citenamefont {Ruan}, \citenamefont {Plutz}, \citenamefont {Belmont},\ and\ \citenamefont {Gratton}}]{Levi:2005}%
  \BibitemOpen
  \bibfield  {author} {\bibinfo {author} {\bibfnamefont {V.}~\bibnamefont {Levi}}, \bibinfo {author} {\bibfnamefont {Q.}~\bibnamefont {Ruan}}, \bibinfo {author} {\bibfnamefont {M.}~\bibnamefont {Plutz}}, \bibinfo {author} {\bibfnamefont {A.~S.}\ \bibnamefont {Belmont}},\ and\ \bibinfo {author} {\bibfnamefont {E.}~\bibnamefont {Gratton}},\ }\bibfield  {title} {\bibinfo {title} {Chromatin dynamics in interphase cells revealed by tracking in a two-photon excitation microscope},\ }\href {https://doi.org/https://doi.org/10.1529/biophysj.105.066670} {\bibfield  {journal} {\bibinfo  {journal} {Biophys. J.}\ }\textbf {\bibinfo {volume} {89}},\ \bibinfo {pages} {4275} (\bibinfo {year} {2005})}\BibitemShut {NoStop}%
\bibitem [{\citenamefont {Zidovska}\ \emph {et~al.}(2013)\citenamefont {Zidovska}, \citenamefont {Weitz},\ and\ \citenamefont {Mitchison}}]{Zidovska:2013}%
  \BibitemOpen
  \bibfield  {author} {\bibinfo {author} {\bibfnamefont {A.}~\bibnamefont {Zidovska}}, \bibinfo {author} {\bibfnamefont {D.~A.}\ \bibnamefont {Weitz}},\ and\ \bibinfo {author} {\bibfnamefont {T.~J.}\ \bibnamefont {Mitchison}},\ }\bibfield  {title} {\bibinfo {title} {Micron-scale coherence in interphase chromatin dynamics},\ }\href {https://doi.org/10.1073/pnas.1220313110} {\bibfield  {journal} {\bibinfo  {journal} {Proc. Natl. Acad. Sci.}\ }\textbf {\bibinfo {volume} {110}},\ \bibinfo {pages} {15555} (\bibinfo {year} {2013})}\BibitemShut {NoStop}%
\bibitem [{\citenamefont {Saintillan}\ \emph {et~al.}(2018)\citenamefont {Saintillan}, \citenamefont {Shelley},\ and\ \citenamefont {Zidovska}}]{Saintillan:2018}%
  \BibitemOpen
  \bibfield  {author} {\bibinfo {author} {\bibfnamefont {D.}~\bibnamefont {Saintillan}}, \bibinfo {author} {\bibfnamefont {M.~J.}\ \bibnamefont {Shelley}},\ and\ \bibinfo {author} {\bibfnamefont {A.}~\bibnamefont {Zidovska}},\ }\bibfield  {title} {\bibinfo {title} {Extensile motor activity drives coherent motions in a model of interphase chromatin},\ }\href {https://doi.org/10.1073/pnas.1807073115} {\bibfield  {journal} {\bibinfo  {journal} {Proc. Natl. Acad. Sci.}\ }\textbf {\bibinfo {volume} {115}},\ \bibinfo {pages} {11442} (\bibinfo {year} {2018})}\BibitemShut {NoStop}%
\bibitem [{\citenamefont {Deblais}\ \emph {et~al.}(2020{\natexlab{a}})\citenamefont {Deblais}, \citenamefont {Woutersen},\ and\ \citenamefont {Bonn}}]{deblais_rheology_2020}%
  \BibitemOpen
  \bibfield  {author} {\bibinfo {author} {\bibfnamefont {A.}~\bibnamefont {Deblais}}, \bibinfo {author} {\bibfnamefont {S.}~\bibnamefont {Woutersen}},\ and\ \bibinfo {author} {\bibfnamefont {D.}~\bibnamefont {Bonn}},\ }\bibfield  {title} {\bibinfo {title} {Rheology of entangled active polymer-like {T. Tubifex} {Worms}},\ }\href {https://doi.org/10.1103/PhysRevLett.124.188002} {\bibfield  {journal} {\bibinfo  {journal} {Phys. Rev. Lett.}\ }\textbf {\bibinfo {volume} {124}},\ \bibinfo {pages} {188002} (\bibinfo {year} {2020}{\natexlab{a}})}\BibitemShut {NoStop}%
\bibitem [{\citenamefont {Deblais}\ \emph {et~al.}(2020{\natexlab{b}})\citenamefont {Deblais}, \citenamefont {Maggs}, \citenamefont {Bonn},\ and\ \citenamefont {Woutersen}}]{deblais_phase_2020}%
  \BibitemOpen
  \bibfield  {author} {\bibinfo {author} {\bibfnamefont {A.}~\bibnamefont {Deblais}}, \bibinfo {author} {\bibfnamefont {A.~C.}\ \bibnamefont {Maggs}}, \bibinfo {author} {\bibfnamefont {D.}~\bibnamefont {Bonn}},\ and\ \bibinfo {author} {\bibfnamefont {S.}~\bibnamefont {Woutersen}},\ }\bibfield  {title} {\bibinfo {title} {Phase separation by entanglement of active polymerlike worms},\ }\href {https://doi.org/10.1103/PhysRevLett.124.208006} {\bibfield  {journal} {\bibinfo  {journal} {Phys. Rev. Lett.}\ }\textbf {\bibinfo {volume} {124}},\ \bibinfo {pages} {208006} (\bibinfo {year} {2020}{\natexlab{b}})}\BibitemShut {NoStop}%
\bibitem [{\citenamefont {Ozkan-Aydin}\ \emph {et~al.}(2021)\citenamefont {Ozkan-Aydin}, \citenamefont {Goldman},\ and\ \citenamefont {Bhamla}}]{ozkan-aydin_collective_2021}%
  \BibitemOpen
  \bibfield  {author} {\bibinfo {author} {\bibfnamefont {Y.}~\bibnamefont {Ozkan-Aydin}}, \bibinfo {author} {\bibfnamefont {D.~I.}\ \bibnamefont {Goldman}},\ and\ \bibinfo {author} {\bibfnamefont {M.~S.}\ \bibnamefont {Bhamla}},\ }\bibfield  {title} {\bibinfo {title} {Collective dynamics in entangled worm and robot blobs},\ }\href {https://doi.org/10.1073/pnas.2010542118} {\bibfield  {journal} {\bibinfo  {journal} {Proc. Natl. Acad. Sci.}\ }\textbf {\bibinfo {volume} {118}},\ \bibinfo {pages} {e2010542118} (\bibinfo {year} {2021})}\BibitemShut {NoStop}%
\bibitem [{\citenamefont {Patil}\ \emph {et~al.}(2023)\citenamefont {Patil}, \citenamefont {Tuazon}, \citenamefont {Kaufman}, \citenamefont {Chakrabortty}, \citenamefont {Qin}, \citenamefont {Dunkel},\ and\ \citenamefont {Bhamla}}]{Patil:2023science}%
  \BibitemOpen
  \bibfield  {author} {\bibinfo {author} {\bibfnamefont {V.~P.}\ \bibnamefont {Patil}}, \bibinfo {author} {\bibfnamefont {H.}~\bibnamefont {Tuazon}}, \bibinfo {author} {\bibfnamefont {E.}~\bibnamefont {Kaufman}}, \bibinfo {author} {\bibfnamefont {T.}~\bibnamefont {Chakrabortty}}, \bibinfo {author} {\bibfnamefont {D.}~\bibnamefont {Qin}}, \bibinfo {author} {\bibfnamefont {J.}~\bibnamefont {Dunkel}},\ and\ \bibinfo {author} {\bibfnamefont {M.~S.}\ \bibnamefont {Bhamla}},\ }\bibfield  {title} {\bibinfo {title} {Ultrafast reversible self-assembly of living tangled matter},\ }\href {https://doi.org/10.1126/science.ade7759} {\bibfield  {journal} {\bibinfo  {journal} {Science}\ }\textbf {\bibinfo {volume} {380}},\ \bibinfo {pages} {392} (\bibinfo {year} {2023})}\BibitemShut {NoStop}%
\bibitem [{\citenamefont {Yardimci}\ \emph {et~al.}(2023)\citenamefont {Yardimci}, \citenamefont {Gibaud}, \citenamefont {Schwenger}, \citenamefont {Sartucci}, \citenamefont {Olmsted}, \citenamefont {Urbach},\ and\ \citenamefont {Dogic}}]{Yardimci2023bonded}%
  \BibitemOpen
  \bibfield  {author} {\bibinfo {author} {\bibfnamefont {S.}~\bibnamefont {Yardimci}}, \bibinfo {author} {\bibfnamefont {T.}~\bibnamefont {Gibaud}}, \bibinfo {author} {\bibfnamefont {W.}~\bibnamefont {Schwenger}}, \bibinfo {author} {\bibfnamefont {M.~R.}\ \bibnamefont {Sartucci}}, \bibinfo {author} {\bibfnamefont {P.~D.}\ \bibnamefont {Olmsted}}, \bibinfo {author} {\bibfnamefont {J.~S.}\ \bibnamefont {Urbach}},\ and\ \bibinfo {author} {\bibfnamefont {Z.}~\bibnamefont {Dogic}},\ }\bibfield  {title} {\bibinfo {title} {Bonded straight and helical flagellar filaments form ultra-low-density glasses},\ }\href {https://doi.org/10.1073/pnas.2215766120} {\bibfield  {journal} {\bibinfo  {journal} {Proc. Natl. Acad. Sci.}\ }\textbf {\bibinfo {volume} {120}},\ \bibinfo {pages} {e2215766120} (\bibinfo {year} {2023})}\BibitemShut {NoStop}%
\bibitem [{\citenamefont {Deblais}\ \emph {et~al.}(2023)\citenamefont {Deblais}, \citenamefont {Prathyusha}, \citenamefont {Sinaasappel}, \citenamefont {Tuazon}, \citenamefont {Tiwari}, \citenamefont {Patil},\ and\ \citenamefont {Bhamla}}]{Deblais2023worm}%
  \BibitemOpen
  \bibfield  {author} {\bibinfo {author} {\bibfnamefont {A.}~\bibnamefont {Deblais}}, \bibinfo {author} {\bibfnamefont {K.}~\bibnamefont {Prathyusha}}, \bibinfo {author} {\bibfnamefont {R.}~\bibnamefont {Sinaasappel}}, \bibinfo {author} {\bibfnamefont {H.}~\bibnamefont {Tuazon}}, \bibinfo {author} {\bibfnamefont {I.}~\bibnamefont {Tiwari}}, \bibinfo {author} {\bibfnamefont {V.~P.}\ \bibnamefont {Patil}},\ and\ \bibinfo {author} {\bibfnamefont {M.~S.}\ \bibnamefont {Bhamla}},\ }\bibfield  {title} {\bibinfo {title} {Worm blobs as entangled living polymers: From topological active matter to flexible soft robot collectives},\ } \href{https://doi.org/10.1039/D3SM00542A} {\bibfield  {journal} {\bibinfo  {journal} {Soft Matter}\ } \textbf {\bibinfo {volume} {19}},\ \bibinfo {pages} {7057-7069} (\bibinfo {year} {2023})}\BibitemShut {NoStop}%
\bibitem [{\citenamefont {Wingstrand}\ \emph {et~al.}(2015)\citenamefont {Wingstrand}, \citenamefont {Alvarez}, \citenamefont {Huang},\ and\ \citenamefont {Hassager}}]{Wingstrand:2015}%
  \BibitemOpen
  \bibfield  {author} {\bibinfo {author} {\bibfnamefont {S.~L.}\ \bibnamefont {Wingstrand}}, \bibinfo {author} {\bibfnamefont {N.~J.}\ \bibnamefont {Alvarez}}, \bibinfo {author} {\bibfnamefont {Q.}~\bibnamefont {Huang}},\ and\ \bibinfo {author} {\bibfnamefont {O.}~\bibnamefont {Hassager}},\ }\bibfield  {title} {\bibinfo {title} {Linear and nonlinear universality in the rheology of polymer melts and solutions},\ }\href {https://doi.org/10.1103/PhysRevLett.115.078302} {\bibfield  {journal} {\bibinfo  {journal} {Phys. Rev. Lett.}\ }\textbf {\bibinfo {volume} {115}},\ \bibinfo {pages} {078302} (\bibinfo {year} {2015})}\BibitemShut {NoStop}%
\bibitem [{\citenamefont {Smrek}\ \emph {et~al.}(2021)\citenamefont {Smrek}, \citenamefont {Garamella}, \citenamefont {Robertson-Anderson},\ and\ \citenamefont {Michieletto}}]{Smrek:2021}%
  \BibitemOpen
  \bibfield  {author} {\bibinfo {author} {\bibfnamefont {J.}~\bibnamefont {Smrek}}, \bibinfo {author} {\bibfnamefont {J.}~\bibnamefont {Garamella}}, \bibinfo {author} {\bibfnamefont {R.}~\bibnamefont {Robertson-Anderson}},\ and\ \bibinfo {author} {\bibfnamefont {D.}~\bibnamefont {Michieletto}},\ }\bibfield  {title} {\bibinfo {title} {Topological tuning of {DNA} mobility in entangled solutions of supercoiled plasmids},\ }\href {https://doi.org/10.1126/sciadv.abf9260} {\bibfield  {journal} {\bibinfo  {journal} {Sci. Adv.}\ }\textbf {\bibinfo {volume} {7}},\ \bibinfo {pages} {eabf9260} (\bibinfo {year} {2021})}\BibitemShut {NoStop}%
\bibitem [{\citenamefont {Kapnistos}\ \emph {et~al.}(2008)\citenamefont {Kapnistos}, \citenamefont {Lang}, \citenamefont {Vlassopoulos}, \citenamefont {Pyckhout-Hintzen}, \citenamefont {Richter}, \citenamefont {Cho}, \citenamefont {Chang},\ and\ \citenamefont {Rubinstein}}]{Kapnistos:2008}%
  \BibitemOpen
  \bibfield  {author} {\bibinfo {author} {\bibfnamefont {M.}~\bibnamefont {Kapnistos}}, \bibinfo {author} {\bibfnamefont {M.}~\bibnamefont {Lang}}, \bibinfo {author} {\bibfnamefont {D.}~\bibnamefont {Vlassopoulos}}, \bibinfo {author} {\bibfnamefont {W.}~\bibnamefont {Pyckhout-Hintzen}}, \bibinfo {author} {\bibfnamefont {D.}~\bibnamefont {Richter}}, \bibinfo {author} {\bibfnamefont {D.}~\bibnamefont {Cho}}, \bibinfo {author} {\bibfnamefont {T.}~\bibnamefont {Chang}},\ and\ \bibinfo {author} {\bibfnamefont {M.}~\bibnamefont {Rubinstein}},\ }\bibfield  {title} {\bibinfo {title} {Unexpected power-law stress relaxation of entangled ring polymers},\ }\href {https://doi.org/10.1038/nmat2292} {\bibfield  {journal} {\bibinfo  {journal} {Nat. Mater.}\ }\textbf {\bibinfo {volume} {7}},\ \bibinfo {pages} {997} (\bibinfo {year} {2008})}\BibitemShut {NoStop}%
\bibitem [{\citenamefont {Doi}\ and\ \citenamefont {Edwards}(1986)}]{doi_theory_1986}%
  \BibitemOpen
  \bibfield  {author} {\bibinfo {author} {\bibfnamefont {M.}~\bibnamefont {Doi}}\ and\ \bibinfo {author} {\bibfnamefont {S.~F.}\ \bibnamefont {Edwards}},\ }\bibfield  {title} {{\selectlanguage {english}\bibinfo {title} {The theory of polymer dynamics}},\ }\bibfield  {journal} {\bibinfo  {journal} {The Clarendon Press, Oxford University Press}\ }\href {https://global.oup.com/academic/product/the-theory-of-polymer-dynamics-9780198520337?cc=de&lang=en&#} (\bibinfo {year} {1986})\BibitemShut {NoStop}%
\bibitem [{\citenamefont {Rubinstein}\ and\ \citenamefont {Colby}(2003)}]{rubinstein_polymer_2003}%
  \BibitemOpen
  \bibfield  {author} {\bibinfo {author} {\bibfnamefont {M.}~\bibnamefont {Rubinstein}}\ and\ \bibinfo {author} {\bibfnamefont {R.~H.}\ \bibnamefont {Colby}},\ }\bibfield  {title} {{\selectlanguage {english}\bibinfo {title} {Polymer physics}},\ }\bibfield  {journal} {\bibinfo  {journal} {Oxford University Press}\ }\href {https://global.oup.com/academic/product/polymer-physics-9780198520597?cc=de&lang=en&} (\bibinfo {year} {2003})\BibitemShut {NoStop}%
\bibitem [{\citenamefont {De~Gennes}(1971)}]{deGennes:1971}%
  \BibitemOpen
  \bibfield  {author} {\bibinfo {author} {\bibfnamefont {P.-G.}\ \bibnamefont {De~Gennes}},\ }\bibfield  {title} {\bibinfo {title} {Reptation of a polymer chain in the presence of fixed obstacles},\ }\href {https://doi.org/10.1063/1.1675789} {\bibfield  {journal} {\bibinfo  {journal} {J. Chem. Phys.}\ }\textbf {\bibinfo {volume} {55}},\ \bibinfo {pages} {572} (\bibinfo {year} {1971})}\BibitemShut {NoStop}%
\bibitem [{\citenamefont {Everaers}\ \emph {et~al.}(2004)\citenamefont {Everaers}, \citenamefont {Sukumaran}, \citenamefont {Grest}, \citenamefont {Svaneborg}, \citenamefont {Sivasubramanian},\ and\ \citenamefont {Kremer}}]{everaers_rheology_2004}%
  \BibitemOpen
  \bibfield  {author} {\bibinfo {author} {\bibfnamefont {R.}~\bibnamefont {Everaers}}, \bibinfo {author} {\bibfnamefont {S.~K.}\ \bibnamefont {Sukumaran}}, \bibinfo {author} {\bibfnamefont {G.~S.}\ \bibnamefont {Grest}}, \bibinfo {author} {\bibfnamefont {C.}~\bibnamefont {Svaneborg}}, \bibinfo {author} {\bibfnamefont {A.}~\bibnamefont {Sivasubramanian}},\ and\ \bibinfo {author} {\bibfnamefont {K.}~\bibnamefont {Kremer}},\ }\bibfield  {title} {\bibinfo {title} {Rheology and {Microscopic} {Topology} of {Entangled} {Polymeric} {Liquids}},\ }\href {https://doi.org/10.1126/science.1091215} {\bibfield  {journal} {\bibinfo  {journal} {Science}\ }\textbf {\bibinfo {volume} {303}},\ \bibinfo {pages} {823} (\bibinfo {year} {2004})}\BibitemShut {NoStop}%
\bibitem [{\citenamefont {Hoy}\ and\ \citenamefont {Kr\"oger}(2020)}]{Hoy:2020PRL}%
  \BibitemOpen
  \bibfield  {author} {\bibinfo {author} {\bibfnamefont {R.~S.}\ \bibnamefont {Hoy}}\ and\ \bibinfo {author} {\bibfnamefont {M.}~\bibnamefont {Kr\"oger}},\ }\bibfield  {title} {\bibinfo {title} {Unified analytic expressions for the entanglement length, tube diameter, and plateau modulus of polymer melts},\ }\href {https://doi.org/10.1103/PhysRevLett.124.147801} {\bibfield  {journal} {\bibinfo  {journal} {Phys. Rev. Lett.}\ }\textbf {\bibinfo {volume} {124}},\ \bibinfo {pages} {147801} (\bibinfo {year} {2020})}\BibitemShut {NoStop}%
\bibitem [{\citenamefont {Hou}\ \emph {et~al.}(2010)\citenamefont {Hou}, \citenamefont {Svaneborg}, \citenamefont {Everaers},\ and\ \citenamefont {Grest}}]{hou_stress_2010}%
  \BibitemOpen
  \bibfield  {author} {\bibinfo {author} {\bibfnamefont {J.-X.}\ \bibnamefont {Hou}}, \bibinfo {author} {\bibfnamefont {C.}~\bibnamefont {Svaneborg}}, \bibinfo {author} {\bibfnamefont {R.}~\bibnamefont {Everaers}},\ and\ \bibinfo {author} {\bibfnamefont {G.~S.}\ \bibnamefont {Grest}},\ }\bibfield  {title} {\bibinfo {title} {Stress relaxation in entangled polymer melts},\ }\href {https://doi.org/10.1103/PhysRevLett.105.068301} {\bibfield  {journal} {\bibinfo  {journal} {Phys. Rev. Lett.}\ }\textbf {\bibinfo {volume} {105}},\ \bibinfo {pages} {068301} (\bibinfo {year} {2010})}\BibitemShut {NoStop}%
\bibitem [{\citenamefont {Kremer}\ and\ \citenamefont {Grest}(1990)}]{Kremer:1990model}%
  \BibitemOpen
  \bibfield  {author} {\bibinfo {author} {\bibfnamefont {K.}~\bibnamefont {Kremer}}\ and\ \bibinfo {author} {\bibfnamefont {G.~S.}\ \bibnamefont {Grest}},\ }\bibfield  {title} {\bibinfo {title} {Dynamics of entangled linear polymer melts: A molecular-dynamics simulation},\ }\href {https://doi.org/10.1063/1.458541} {\bibfield  {journal} {\bibinfo  {journal} {J. Chem. Phys.}\ }\textbf {\bibinfo {volume} {92}},\ \bibinfo {pages} {5057} (\bibinfo {year} {1990})}\BibitemShut {NoStop}%
\bibitem [{\citenamefont {Weeks}\ \emph {et~al.}(1971)\citenamefont {Weeks}, \citenamefont {Chandler},\ and\ \citenamefont {Andersen}}]{Weeks:1971wca}%
  \BibitemOpen
  \bibfield  {author} {\bibinfo {author} {\bibfnamefont {J.~D.}\ \bibnamefont {Weeks}}, \bibinfo {author} {\bibfnamefont {D.}~\bibnamefont {Chandler}},\ and\ \bibinfo {author} {\bibfnamefont {H.~C.}\ \bibnamefont {Andersen}},\ }\bibfield  {title} {\bibinfo {title} {Role of repulsive forces in determining the equilibrium structure of simple liquids},\ }\href {https://doi.org/10.1063/1.1674820} {\bibfield  {journal} {\bibinfo  {journal} {J. Chem. Phys.}\ }\textbf {\bibinfo {volume} {54}},\ \bibinfo {pages} {5237} (\bibinfo {year} {1971})}\BibitemShut {NoStop}%
\bibitem [{\citenamefont {Kurzthaler}\ \emph {et~al.}(2021)\citenamefont {Kurzthaler}, \citenamefont {Mandal}, \citenamefont {Bhattacharjee}, \citenamefont {Löwen}, \citenamefont {Datta},\ and\ \citenamefont {Stone}}]{kurzthaler_geometric_2021}%
  \BibitemOpen
  \bibfield  {author} {\bibinfo {author} {\bibfnamefont {C.}~\bibnamefont {Kurzthaler}}, \bibinfo {author} {\bibfnamefont {S.}~\bibnamefont {Mandal}}, \bibinfo {author} {\bibfnamefont {T.}~\bibnamefont {Bhattacharjee}}, \bibinfo {author} {\bibfnamefont {H.}~\bibnamefont {Löwen}}, \bibinfo {author} {\bibfnamefont {S.~S.}\ \bibnamefont {Datta}},\ and\ \bibinfo {author} {\bibfnamefont {H.~A.}\ \bibnamefont {Stone}},\ }\bibfield  {title} {{\selectlanguage {english}\bibinfo {title} {A geometric criterion for the optimal spreading of active polymers in porous media}},\ }\href {https://doi.org/10.1038/s41467-021-26942-0} {\bibfield  {journal} {\bibinfo  {journal} {Nat. Commun.}\ }\textbf {\bibinfo {volume} {12}},\ \bibinfo {pages} {7088} (\bibinfo {year} {2021})}\BibitemShut {NoStop}%
\bibitem [{\citenamefont {Bianco}\ \emph {et~al.}(2018)\citenamefont {Bianco}, \citenamefont {Locatelli},\ and\ \citenamefont {Malgaretti}}]{Bianco:2018}%
  \BibitemOpen
  \bibfield  {author} {\bibinfo {author} {\bibfnamefont {V.}~\bibnamefont {Bianco}}, \bibinfo {author} {\bibfnamefont {E.}~\bibnamefont {Locatelli}},\ and\ \bibinfo {author} {\bibfnamefont {P.}~\bibnamefont {Malgaretti}},\ }\bibfield  {title} {\bibinfo {title} {Globulelike conformation and enhanced diffusion of active polymers},\ }\href {https://doi.org/10.1103/PhysRevLett.121.217802} {\bibfield  {journal} {\bibinfo  {journal} {Phys. Rev. Lett.}\ }\textbf {\bibinfo {volume} {121}},\ \bibinfo {pages} {217802} (\bibinfo {year} {2018})}\BibitemShut {NoStop}%
\bibitem [{\citenamefont {Prathyusha}\ \emph {et~al.}(2018)\citenamefont {Prathyusha}, \citenamefont {Henkes},\ and\ \citenamefont {Sknepnek}}]{Prathyusha:2018}%
  \BibitemOpen
  \bibfield  {author} {\bibinfo {author} {\bibfnamefont {K.~R.}\ \bibnamefont {Prathyusha}}, \bibinfo {author} {\bibfnamefont {S.}~\bibnamefont {Henkes}},\ and\ \bibinfo {author} {\bibfnamefont {R.}~\bibnamefont {Sknepnek}},\ }\bibfield  {title} {\bibinfo {title} {Dynamically generated patterns in dense suspensions of active filaments},\ }\href {https://doi.org/10.1103/PhysRevE.97.022606} {\bibfield  {journal} {\bibinfo  {journal} {Phys. Rev. E}\ }\textbf {\bibinfo {volume} {97}},\ \bibinfo {pages} {022606} (\bibinfo {year} {2018})}\BibitemShut {NoStop}%
\bibitem [{\citenamefont {Kr{\"o}ger}\ \emph {et~al.}(2023)\citenamefont {Kr{\"o}ger}, \citenamefont {Dietz}, \citenamefont {Hoy},\ and\ \citenamefont {Luap}}]{kroger_z1_2023}%
  \BibitemOpen
  \bibfield  {author} {\bibinfo {author} {\bibfnamefont {M.}~\bibnamefont {Kr{\"o}ger}}, \bibinfo {author} {\bibfnamefont {J.~D.}\ \bibnamefont {Dietz}}, \bibinfo {author} {\bibfnamefont {R.~S.}\ \bibnamefont {Hoy}},\ and\ \bibinfo {author} {\bibfnamefont {C.}~\bibnamefont {Luap}},\ }\bibfield  {title} {\bibinfo {title} {The z1+ package: Shortest multiple disconnected path for the analysis of entanglements in macromolecular systems},\ }\href {https://doi.org/10.1016/j.cpc.2022.108567} {\bibfield  {journal} {\bibinfo  {journal} {Comput. Phys. Commun.}\ }\textbf {\bibinfo {volume} {283}},\ \bibinfo {pages} {108567} (\bibinfo {year} {2023})}\BibitemShut {NoStop}%
\bibitem [{sup()}]{suppl}%
  \BibitemOpen
  \href@noop {} {}\bibinfo {note} {See Supplemental Material at [URL], for details of the equilibration process, primitive path analysis, and investigation of viscoelasticity at shorter polymer lengths across various P\'eclet numbers.}\BibitemShut {Stop}%
\bibitem [{\citenamefont {Kr{\"o}ger}(2005)}]{kroger_shortest_2005}%
  \BibitemOpen
  \bibfield  {author} {\bibinfo {author} {\bibfnamefont {M.}~\bibnamefont {Kr{\"o}ger}},\ }\bibfield  {title} {\bibinfo {title} {Shortest multiple disconnected path for the analysis of entanglements in two-and three-dimensional polymeric systems},\ }\href {https://doi.org/10.1016/j.cpc.2005.01.020} {\bibfield  {journal} {\bibinfo  {journal} {Comput. Phys. Commun.}\ }\textbf {\bibinfo {volume} {168}},\ \bibinfo {pages} {209} (\bibinfo {year} {2005})}\BibitemShut {NoStop}%
\bibitem [{\citenamefont {Shanbhag}\ and\ \citenamefont {Kr\"oger}(2007)}]{shanbhag_primitive_2007}%
  \BibitemOpen
 \bibfield  {author} {\bibinfo {author} {\bibfnamefont {S.}~\bibnamefont {Shanbhag}}\ and\ \bibinfo {author} {\bibfnamefont {M.}~\bibnamefont {Kr\"oger}},\ }\bibfield  {title} {\bibinfo {title} {Primitive Path Networks Generated by Annealing and Geometrical Methods: Insights into Differences},\ }\href {https://doi.org/10.1021/ma062457k} {\bibfield  {journal} {\bibinfo  {journal} {Macromolecules}\ }\textbf {\bibinfo {volume} {40}},\ \bibinfo {pages} {2897} (\bibinfo {year} {2007})}\BibitemShut {NoStop}%
\bibitem [{\citenamefont {Hoy}\ \emph {et~al.}(2009)\citenamefont {Hoy}, \citenamefont {Foteinopoulou},\ and\ \citenamefont {Kr\"oger}}]{hoy_topological_2009}%
  \BibitemOpen
  \bibfield  {author} {\bibinfo {author} {\bibfnamefont {R.~S.}\ \bibnamefont {Hoy}}, \bibinfo {author} {\bibfnamefont {K.}~\bibnamefont {Foteinopoulou}},\ and\ \bibinfo {author} {\bibfnamefont {M.}~\bibnamefont {Kr\"oger}},\ }\bibfield  {title} {\bibinfo {title} {Topological analysis of polymeric melts: Chain-length effects and fast-converging estimators for entanglement length},\ }\href {https://doi.org/10.1103/PhysRevE.80.031803} {\bibfield  {journal} {\bibinfo  {journal} {Phys. Rev. E}\ }\textbf {\bibinfo {volume} {80}},\ \bibinfo {pages} {031803} (\bibinfo {year} {2009})}\BibitemShut {NoStop}%
\bibitem [{\citenamefont {Karayiannis}\ and\ \citenamefont {Kröger}(2009)}]{karayiannis_combined_2009}%
  \BibitemOpen
  \bibfield  {author} {\bibinfo {author} {\bibfnamefont {N.~C.}\ \bibnamefont {Karayiannis}}\ and\ \bibinfo {author} {\bibfnamefont {M.}~\bibnamefont {Kröger}},\ }\bibfield  {title} {{\selectlanguage {english}\bibinfo {title} {Combined {Molecular} {Algorithms} for the {Generation}, {Equilibration} and {Topological} {Analysis} of {Entangled} {Polymers}: {Methodology} and {Performance}}},\ }\href {https://doi.org/10.3390/ijms10115054} {\bibfield  {journal} {\bibinfo  {journal} {Int. J. Mol. Sci.}\ }\textbf {\bibinfo {volume} {10}},\ \bibinfo {pages} {5054} (\bibinfo {year} {2009})}\BibitemShut {NoStop}%
\bibitem [{\citenamefont {Hsu}\ and\ \citenamefont {Kremer}(2016)}]{hsu_static_2016}%
  \BibitemOpen
  \bibfield  {author} {\bibinfo {author} {\bibfnamefont {H.-P.}\ \bibnamefont {Hsu}}\ and\ \bibinfo {author} {\bibfnamefont {K.}~\bibnamefont {Kremer}},\ }\bibfield  {title} {\bibinfo {title} {Static and dynamic properties of large polymer melts in equilibrium},\ }\href {https://doi.org/10.1063/1.4946033} {\bibfield  {journal} {\bibinfo  {journal} {J. Chem. Phys.}\ }\textbf {\bibinfo {volume} {144}},\ \bibinfo {pages} {154907} (\bibinfo {year} {2016})}\BibitemShut {NoStop}%
\bibitem [{\citenamefont {Keshavarz}\ \emph {et~al.}(2016)\citenamefont {Keshavarz}, \citenamefont {Engelkamp}, \citenamefont {Xu}, \citenamefont {Braeken}, \citenamefont {Otten}, \citenamefont {Uji-i}, \citenamefont {Schwartz}, \citenamefont {Koepf}, \citenamefont {Vananroye}, \citenamefont {Vermant} \emph {et~al.}}]{Keshavarz2016:nanoscale}%
  \BibitemOpen
  \bibfield  {author} {\bibinfo {author} {\bibfnamefont {M.}~\bibnamefont {Keshavarz}}, \bibinfo {author} {\bibfnamefont {H.}~\bibnamefont {Engelkamp}}, \bibinfo {author} {\bibfnamefont {J.}~\bibnamefont {Xu}}, \bibinfo {author} {\bibfnamefont {E.}~\bibnamefont {Braeken}}, \bibinfo {author} {\bibfnamefont {M.~B.}\ \bibnamefont {Otten}}, \bibinfo {author} {\bibfnamefont {H.}~\bibnamefont {Uji-i}}, \bibinfo {author} {\bibfnamefont {E.}~\bibnamefont {Schwartz}}, \bibinfo {author} {\bibfnamefont {M.}~\bibnamefont {Koepf}}, \bibinfo {author} {\bibfnamefont {A.}~\bibnamefont {Vananroye}}, \bibinfo {author} {\bibfnamefont {J.}~\bibnamefont {Vermant}}, \emph {et~al.},\ }\bibfield  {title} {\bibinfo {title} {Nanoscale study of polymer dynamics},\ }\href {https://doi.org/10.1021/acsnano.5b06931} {\bibfield  {journal} {\bibinfo  {journal} {ACS nano}\ }\textbf {\bibinfo {volume} {10}},\ \bibinfo {pages} {1434} (\bibinfo {year} {2016})}\BibitemShut {NoStop}%
\bibitem [{\citenamefont {Han}\ \emph {et~al.}(2021)\citenamefont {Han}, \citenamefont {Fruchart}, \citenamefont {Scheibner}, \citenamefont {Vaikuntanathan}, \citenamefont {De~Pablo},\ and\ \citenamefont {Vitelli}}]{Han_fluctuating:2021}%
  \BibitemOpen
  \bibfield  {author} {\bibinfo {author} {\bibfnamefont {M.}~\bibnamefont {Han}}, \bibinfo {author} {\bibfnamefont {M.}~\bibnamefont {Fruchart}}, \bibinfo {author} {\bibfnamefont {C.}~\bibnamefont {Scheibner}}, \bibinfo {author} {\bibfnamefont {S.}~\bibnamefont {Vaikuntanathan}}, \bibinfo {author} {\bibfnamefont {J.~J.}\ \bibnamefont {De~Pablo}},\ and\ \bibinfo {author} {\bibfnamefont {V.}~\bibnamefont {Vitelli}},\ }\bibfield  {title} {\bibinfo {title} {Fluctuating hydrodynamics of chiral active fluids},\ }\href {https://doi.org/10.1038/s41567-021-01360-7} {\bibfield  {journal} {\bibinfo  {journal} {Nat. Phys.}\ }\textbf {\bibinfo {volume} {17}},\ \bibinfo {pages} {1260} (\bibinfo {year} {2021})}\BibitemShut {NoStop}%
\bibitem [{\citenamefont {Bacanu}\ \emph {et~al.}(2023)\citenamefont {Bacanu}, \citenamefont {Pelletier}, \citenamefont {Jung},\ and\ \citenamefont {Fakhri}}]{Bacanu_activeNanotubes:2023}%
  \BibitemOpen
  \bibfield  {author} {\bibinfo {author} {\bibfnamefont {A.}~\bibnamefont {Bacanu}}, \bibinfo {author} {\bibfnamefont {J.~F.}\ \bibnamefont {Pelletier}}, \bibinfo {author} {\bibfnamefont {Y.}~\bibnamefont {Jung}},\ and\ \bibinfo {author} {\bibfnamefont {N.}~\bibnamefont {Fakhri}},\ }\bibfield  {title} {\bibinfo {title} {Inferring scale-dependent non-equilibrium activity using carbon nanotubes},\ }\href {https://doi.org/10.1038/s41565-023-01395-2} {\bibfield  {journal} {\bibinfo  {journal} {Nat. Nanotechnol.}\ }\textbf {\bibinfo {volume} {18}},\ \bibinfo {pages} {905} (\bibinfo {year} {2023})}\BibitemShut {NoStop}%
\bibitem [{\citenamefont {Yang}\ \emph {et~al.}(2020)\citenamefont {Yang}, \citenamefont {Sprinkle}, \citenamefont {Guo}, \citenamefont {Qian}, \citenamefont {Hua}, \citenamefont {Donev}, \citenamefont {Marr},\ and\ \citenamefont {Wu}}]{Yang_reconfigurable:2020}%
  \BibitemOpen
  \bibfield  {author} {\bibinfo {author} {\bibfnamefont {T.}~\bibnamefont {Yang}}, \bibinfo {author} {\bibfnamefont {B.}~\bibnamefont {Sprinkle}}, \bibinfo {author} {\bibfnamefont {Y.}~\bibnamefont {Guo}}, \bibinfo {author} {\bibfnamefont {J.}~\bibnamefont {Qian}}, \bibinfo {author} {\bibfnamefont {D.}~\bibnamefont {Hua}}, \bibinfo {author} {\bibfnamefont {A.}~\bibnamefont {Donev}}, \bibinfo {author} {\bibfnamefont {D.~W.}\ \bibnamefont {Marr}},\ and\ \bibinfo {author} {\bibfnamefont {N.}~\bibnamefont {Wu}},\ }\bibfield  {title} {\bibinfo {title} {Reconfigurable microbots folded from simple colloidal chains},\ }\href {https://doi.org/10.1073/pnas.2007255117} {\bibfield  {journal} {\bibinfo  {journal} {Proc. Natl. Acad. Sci.}\ }\textbf {\bibinfo {volume} {117}},\ \bibinfo {pages} {18186} (\bibinfo {year} {2020})}\BibitemShut {NoStop}%
\bibitem [{\citenamefont {Van~Ruymbeke}\ and\ \citenamefont {Vlassopoulos}(2022)}]{vanmacromolecular:2022}%
  \BibitemOpen
  \bibfield  {author} {\bibinfo {author} {\bibfnamefont {E.}~\bibnamefont {Van~Ruymbeke}}\ and\ \bibinfo {author} {\bibfnamefont {D.}~\bibnamefont {Vlassopoulos}},\ }\href {https://doi.org/10.1002/9783527815562.mme0034} {\emph {\bibinfo {title} {Macromolecular Rheology}}}\ (\bibinfo  {publisher} {Wiley},\ \bibinfo {year} {2022})\BibitemShut {NoStop}%
\bibitem [{\citenamefont {Hsu}\ and\ \citenamefont {Kremer}(2019)}]{Hsuclustering:2019}%
  \BibitemOpen
  \bibfield  {author} {\bibinfo {author} {\bibfnamefont {H.-P.}\ \bibnamefont {Hsu}}\ and\ \bibinfo {author} {\bibfnamefont {K.}~\bibnamefont {Kremer}},\ }\bibfield  {title} {\bibinfo {title} {Clustering of entanglement points in highly strained polymer melts},\ }\href {https://doi.org/10.1021/acs.macromol.9b01120} {\bibfield  {journal} {\bibinfo  {journal} {Macromolecules}\ }\textbf {\bibinfo {volume} {52}},\ \bibinfo {pages} {6756} (\bibinfo {year} {2019})}\BibitemShut {NoStop}%
\bibitem [{\citenamefont {Wang}\ \emph {et~al.}(2007)\citenamefont {Wang}, \citenamefont {Ravindranath}, \citenamefont {Wang},\ and\ \citenamefont {Boukany}}]{Wang:2007new}%
  \BibitemOpen
  \bibfield  {author} {\bibinfo {author} {\bibfnamefont {S.-Q.}\ \bibnamefont {Wang}}, \bibinfo {author} {\bibfnamefont {S.}~\bibnamefont {Ravindranath}}, \bibinfo {author} {\bibfnamefont {Y.}~\bibnamefont {Wang}},\ and\ \bibinfo {author} {\bibfnamefont {P.}~\bibnamefont {Boukany}},\ }\bibfield  {title} {\bibinfo {title} {New theoretical considerations in polymer rheology: Elastic breakdown of chain entanglement network},\ }\bibfield  {journal} {\bibinfo  {journal} {J. Chem. Phys.}\ }\textbf {\bibinfo {volume} {127}},\ \href {https://doi.org/10.1063/1.2753156} {10.1063/1.2753156} (\bibinfo {year} {2007})\BibitemShut {NoStop}%
\bibitem [{\citenamefont {Cates}\ \emph {et~al.}(1993)\citenamefont {Cates}, \citenamefont {McLeish},\ and\ \citenamefont {Marrucci}}]{Cates1993:rheology}%
  \BibitemOpen
  \bibfield  {author} {\bibinfo {author} {\bibfnamefont {M.}~\bibnamefont {Cates}}, \bibinfo {author} {\bibfnamefont {T.}~\bibnamefont {McLeish}},\ and\ \bibinfo {author} {\bibfnamefont {G.}~\bibnamefont {Marrucci}},\ }\bibfield  {title} {\bibinfo {title} {The rheology of entangled polymers at very high shear rates},\ }\href {https://doi.org/10.1209/0295-5075/21/4/012} {\bibfield  {journal} {\bibinfo  {journal} {EPL}\ }\textbf {\bibinfo {volume} {21}},\ \bibinfo {pages} {451} (\bibinfo {year} {1993})}\BibitemShut {NoStop}%
\bibitem [{\citenamefont {O’Connor}\ \emph {et~al.}(2019)\citenamefont {O’Connor}, \citenamefont {Hopkins},\ and\ \citenamefont {Robbins}}]{OConnorstress:2019}%
  \BibitemOpen
  \bibfield  {author} {\bibinfo {author} {\bibfnamefont {T.~C.}\ \bibnamefont {O’Connor}}, \bibinfo {author} {\bibfnamefont {A.}~\bibnamefont {Hopkins}},\ and\ \bibinfo {author} {\bibfnamefont {M.~O.}\ \bibnamefont {Robbins}},\ }\bibfield  {title} {\bibinfo {title} {Stress relaxation in highly oriented melts of entangled polymers},\ }\href {https://doi.org/10.1021/acs.macromol.9b01161} {\bibfield  {journal} {\bibinfo  {journal} {Macromolecules}\ }\textbf {\bibinfo {volume} {52}},\ \bibinfo {pages} {8540} (\bibinfo {year} {2019})}\BibitemShut {NoStop}%
\bibitem [{\citenamefont {Banerjee}\ \emph {et~al.}(2020)\citenamefont {Banerjee}, \citenamefont {Gardel},\ and\ \citenamefont {Schwarz}}]{Banerjee:2020actin}%
  \BibitemOpen
  \bibfield  {author} {\bibinfo {author} {\bibfnamefont {S.}~\bibnamefont {Banerjee}}, \bibinfo {author} {\bibfnamefont {M.~L.}\ \bibnamefont {Gardel}},\ and\ \bibinfo {author} {\bibfnamefont {U.~S.}\ \bibnamefont {Schwarz}},\ }\bibfield  {title} {\bibinfo {title} {The actin cytoskeleton as an active adaptive material},\ }\href {https://doi.org/10.1146/annurev-conmatphys-031218-013231} {\bibfield  {journal} {\bibinfo  {journal} {Annu. Rev. Condens. Matter Phys.}\ }\textbf {\bibinfo {volume} {11}},\ \bibinfo {pages} {421} (\bibinfo {year} {2020})}\BibitemShut {NoStop}%
\bibitem [{\citenamefont {Winkler}\ and\ \citenamefont {Gompper}(2020)}]{winkler_physics_2020}%
  \BibitemOpen
  \bibfield  {author} {\bibinfo {author} {\bibfnamefont {R.~G.}\ \bibnamefont {Winkler}}\ and\ \bibinfo {author} {\bibfnamefont {G.}~\bibnamefont {Gompper}},\ }\bibfield  {title} {\bibinfo {title} {The physics of active polymers and filaments},\ }\href {https://doi.org/10.1063/5.0011466} {\bibfield  {journal} {\bibinfo  {journal} {J. Chem. Phys.}\ }\textbf {\bibinfo {volume} {153}},\ \bibinfo {pages} {040901} (\bibinfo {year} {2020})}\BibitemShut {NoStop}%
\bibitem [{\citenamefont {Eisenstecken}\ \emph {et~al.}(2017)\citenamefont {Eisenstecken}, \citenamefont {Gompper},\ and\ \citenamefont {Winkler}}]{Eisenstecken:2017}%
  \BibitemOpen
  \bibfield  {author} {\bibinfo {author} {\bibfnamefont {T.}~\bibnamefont {Eisenstecken}}, \bibinfo {author} {\bibfnamefont {G.}~\bibnamefont {Gompper}},\ and\ \bibinfo {author} {\bibfnamefont {R.~G.}\ \bibnamefont {Winkler}},\ }\bibfield  {title} {\bibinfo {title} {Internal dynamics of semiflexible polymers with active noise},\ }\href {https://doi.org/10.1063/1.4981012} {\bibfield  {journal} {\bibinfo  {journal} {J. Chem. Phys.}\ }\textbf {\bibinfo {volume} {146}},\ \bibinfo {pages} {154903} (\bibinfo {year} {2017})}\BibitemShut {NoStop}%
\bibitem [{\citenamefont {Eisenstecken}\ and\ \citenamefont {Winkler}(2022)}]{Eisenstecken:2022path}%
  \BibitemOpen
  \bibfield  {author} {\bibinfo {author} {\bibfnamefont {T.}~\bibnamefont {Eisenstecken}}\ and\ \bibinfo {author} {\bibfnamefont {R.~G.}\ \bibnamefont {Winkler}},\ }\bibfield  {title} {\bibinfo {title} {Path integral description of semiflexible active {Brownian} polymers},\ }\bibfield  {journal} {\bibinfo  {journal} {J. Chem. Phys.}\ }\textbf {\bibinfo {volume} {156}},\ \href {https://doi.org/10.1063/5.0081020} {10.1063/5.0081020} (\bibinfo {year} {2022})\BibitemShut {NoStop}%
\bibitem [{\citenamefont {Harnau}\ \emph {et~al.}(1996)\citenamefont {Harnau}, \citenamefont {Winkler},\ and\ \citenamefont {Reineker}}]{Harnau:1996dynamic}%
  \BibitemOpen
  \bibfield  {author} {\bibinfo {author} {\bibfnamefont {L.}~\bibnamefont {Harnau}}, \bibinfo {author} {\bibfnamefont {R.~G.}\ \bibnamefont {Winkler}},\ and\ \bibinfo {author} {\bibfnamefont {P.}~\bibnamefont {Reineker}},\ }\bibfield  {title} {\bibinfo {title} {Dynamic structure factor of semiflexible macromolecules in dilute solution},\ }\href {https://doi.org/10.1063/1.471297} {\bibfield  {journal} {\bibinfo  {journal} {J. Chem. Phys.}\ }\textbf {\bibinfo {volume} {104}},\ \bibinfo {pages} {6355} (\bibinfo {year} {1996})}\BibitemShut {NoStop}%
\bibitem [{\citenamefont {Kroy}\ and\ \citenamefont {Frey}(1996)}]{Kroy:1996}%
  \BibitemOpen
  \bibfield  {author} {\bibinfo {author} {\bibfnamefont {K.}~\bibnamefont {Kroy}}\ and\ \bibinfo {author} {\bibfnamefont {E.}~\bibnamefont {Frey}},\ }\bibfield  {title} {\bibinfo {title} {Force-extension relation and plateau modulus for wormlike chains},\ }\href {https://doi.org/10.1103/PhysRevLett.77.306} {\bibfield  {journal} {\bibinfo  {journal} {Phys. Rev. Lett.}\ }\textbf {\bibinfo {volume} {77}},\ \bibinfo {pages} {306} (\bibinfo {year} {1996})}\BibitemShut {NoStop}%
\bibitem [{\citenamefont {Hallatschek}\ \emph {et~al.}(2005)\citenamefont {Hallatschek}, \citenamefont {Frey},\ and\ \citenamefont {Kroy}}]{Hallatschek:2005}%
  \BibitemOpen
  \bibfield  {author} {\bibinfo {author} {\bibfnamefont {O.}~\bibnamefont {Hallatschek}}, \bibinfo {author} {\bibfnamefont {E.}~\bibnamefont {Frey}},\ and\ \bibinfo {author} {\bibfnamefont {K.}~\bibnamefont {Kroy}},\ }\bibfield  {title} {\bibinfo {title} {Propagation and relaxation of tension in stiff polymers},\ }\href {https://doi.org/10.1103/PhysRevLett.94.077804} {\bibfield  {journal} {\bibinfo  {journal} {Phys. Rev. Lett.}\ }\textbf {\bibinfo {volume} {94}},\ \bibinfo {pages} {077804} (\bibinfo {year} {2005})}\BibitemShut {NoStop}%
\bibitem [{\citenamefont {Lang}\ and\ \citenamefont {Frey}(2018)}]{Lang:2018disentangling}%
  \BibitemOpen
  \bibfield  {author} {\bibinfo {author} {\bibfnamefont {P.}~\bibnamefont {Lang}}\ and\ \bibinfo {author} {\bibfnamefont {E.}~\bibnamefont {Frey}},\ }\bibfield  {title} {\bibinfo {title} {Disentangling entanglements in biopolymer solutions},\ }\href {https://doi.org/10.1038/s41467-018-02837-5} {\bibfield  {journal} {\bibinfo  {journal} {Nat. Commun.}\ }\textbf {\bibinfo {volume} {9}},\ \bibinfo {pages} {1} (\bibinfo {year} {2018})}\BibitemShut {NoStop}%
\bibitem [{\citenamefont {Mandal}\ \emph {et~al.}(2020)\citenamefont {Mandal}, \citenamefont {Kurzthaler}, \citenamefont {Franosch},\ and\ \citenamefont {Löwen}}]{mandal_crowding-enhanced_2020}%
  \BibitemOpen
  \bibfield  {author} {\bibinfo {author} {\bibfnamefont {S.}~\bibnamefont {Mandal}}, \bibinfo {author} {\bibfnamefont {C.}~\bibnamefont {Kurzthaler}}, \bibinfo {author} {\bibfnamefont {T.}~\bibnamefont {Franosch}},\ and\ \bibinfo {author} {\bibfnamefont {H.}~\bibnamefont {Löwen}},\ }\bibfield  {title} {\bibinfo {title} {Crowding-{Enhanced} {Diffusion}: {An} {Exact} {Theory} for {Highly} {Entangled} {Self}-{Propelled} {Stiff} {Filaments}},\ }\href {https://doi.org/10.1103/PhysRevLett.125.138002} {\bibfield  {journal} {\bibinfo  {journal} {Phys. Rev. Lett.}\ }\textbf {\bibinfo {volume} {125}},\ \bibinfo {pages} {138002} (\bibinfo {year} {2020})}\BibitemShut {NoStop}%
\bibitem [{\citenamefont {Blain}\ and\ \citenamefont {Szostak}(2014)}]{Blain:2014}%
  \BibitemOpen
  \bibfield  {author} {\bibinfo {author} {\bibfnamefont {J.~C.}\ \bibnamefont {Blain}}\ and\ \bibinfo {author} {\bibfnamefont {J.~W.}\ \bibnamefont {Szostak}},\ }\bibfield  {title} {\bibinfo {title} {Progress toward synthetic cells},\ }\href {https://doi.org/10.1146/annurev-biochem-080411-124036} {\bibfield  {journal} {\bibinfo  {journal} {Annu. Rev. Biochem}\ }\textbf {\bibinfo {volume} {83}},\ \bibinfo {pages} {615} (\bibinfo {year} {2014})}\BibitemShut {NoStop}%
\bibitem [{\citenamefont {Ganzinger}\ and\ \citenamefont {Schwille}(2019)}]{Ganzinger:2019}%
  \BibitemOpen
  \bibfield  {author} {\bibinfo {author} {\bibfnamefont {K.~A.}\ \bibnamefont {Ganzinger}}\ and\ \bibinfo {author} {\bibfnamefont {P.}~\bibnamefont {Schwille}},\ }\bibfield  {title} {\bibinfo {title} {More from less--bottom-up reconstitution of cell biology},\ }\href {https://doi.org/10.1242/jcs.227488} {\bibfield  {journal} {\bibinfo  {journal} {J. Cell Sci.}\ }\textbf {\bibinfo {volume} {132}},\ \bibinfo {pages} {jcs227488} (\bibinfo {year} {2019})}\BibitemShut {NoStop}%
\end{thebibliography}

\ifarXiv
    

\section*{Supplementary material}

# Supplemental Material: Giant Activity-Induced Stress Plateau in Entangled Polymer Solutions

Davide Breoni,[1] Christina Kurzthaler,[2] Benno Liebchen,[3] Hartmut Löwen,[1] and Suvendu Mandal[3, *]

[1] *Institut für Theoretische Physik II: Weiche Materie, Heinrich Heine-Universität Düsseldorf, Universitätsstraße 1, 40225 Düsseldorf, Germany*
[2] *Max Planck Institute for the Physics of Complex Systems, Nöhnitzer Straße 38, 01187 Dresden, Germany*
[3] *Technische Universität Darmstadt, Karolinenplatz 5, 64289 Darmstadt, Germany*

(Dated: September 29, 2023)

### System equilibration

A well-known challenge in the field of entangled polymer physics has been the excessively long timescales required to reach equilibrium, with relaxation times scaling as the cube of the polymer length, i.e., $\sim L^3$. To circumvent the need for exceedingly lengthy simulations, we employ a highly efficient approach known as the double-bridging hybrid (DBH) bond-swapping algorithm, in conjunction with core softening techniques as outlined in Dietz et al.'s work [1]. The DBH algorithm operates by executing Monte Carlo (MC) moves to swap bonds and angles within the context of molecular dynamics simulations [Fig. 1]. Notably, this technique allows for the exchange of entire strands of polymers, a capability not available in standard molecular dynamics simulations. As a result, it substantially reduces the relaxation time, transitioning from the daunting $\sim L^3$ scaling to a much more manageable $\sim L$.

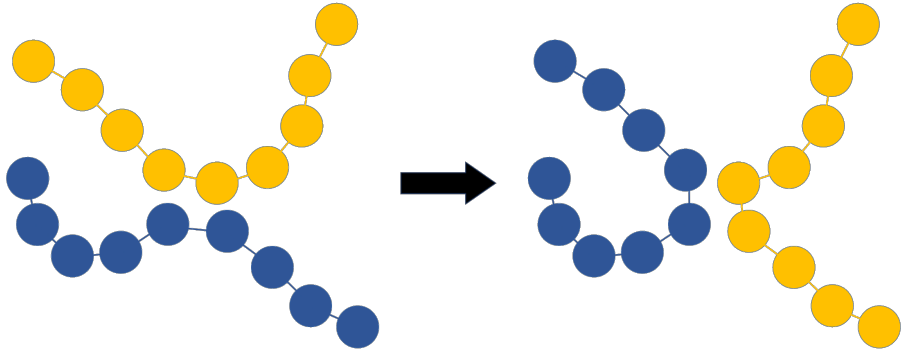

Figure 1. Representation of a double-bridging hybrid Monte Carlo move, demonstrating the exchange of bonds between separate polymer chains.

### Primitive path analysis and topology

To explore the system's topology, we employ the Z1+ algorithm developed by M. Kröger [2]. The Z1+ algorithm iteratively simplifies the initial polymer configuration based on entanglement point positions, thus revealing the essential topological structure of the primitive paths [see Fig. 2(a)]. It begins by examining sets of three consecutive nodes along each polymer, initially defined by monomer positions. It evaluates the area enclosed by the triangle formed by these nodes, accounting for potential obstacles defined by intersecting paths. After multiple iterations, when further area reduction becomes unattainable, the resulting nodes represent the system's topological entanglement points. The average number of entanglement points, denoted as $Z$, is calculated as the mean number of nodes per path, while $L_{pp}$ signifies the average path length. Figure 2(b) presents the final primitive path of a tracer polymer and its neighboring paths obtained using the Z1+ algorithm from a simulation configuration.

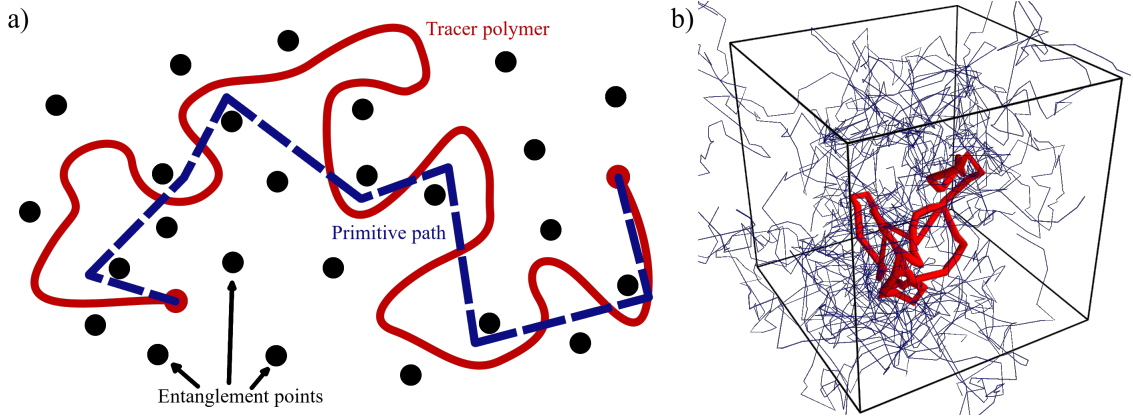

Figure 2. (a) Schematic representation illustrating the operation of the Z1+ algorithm. The primitive path (blue) relative to a tracer polymer (red) is depicted, with entanglement points (black) representing obstacles posed by other polymers. (b) Snapshot from a simulation displaying the primitive paths of a tracer polymer (red) and all of its neighboring polymers (blue). This configuration corresponds to Pe = 0 and $L = 1450\sigma$.

**Polymer conformation and entanglement length**

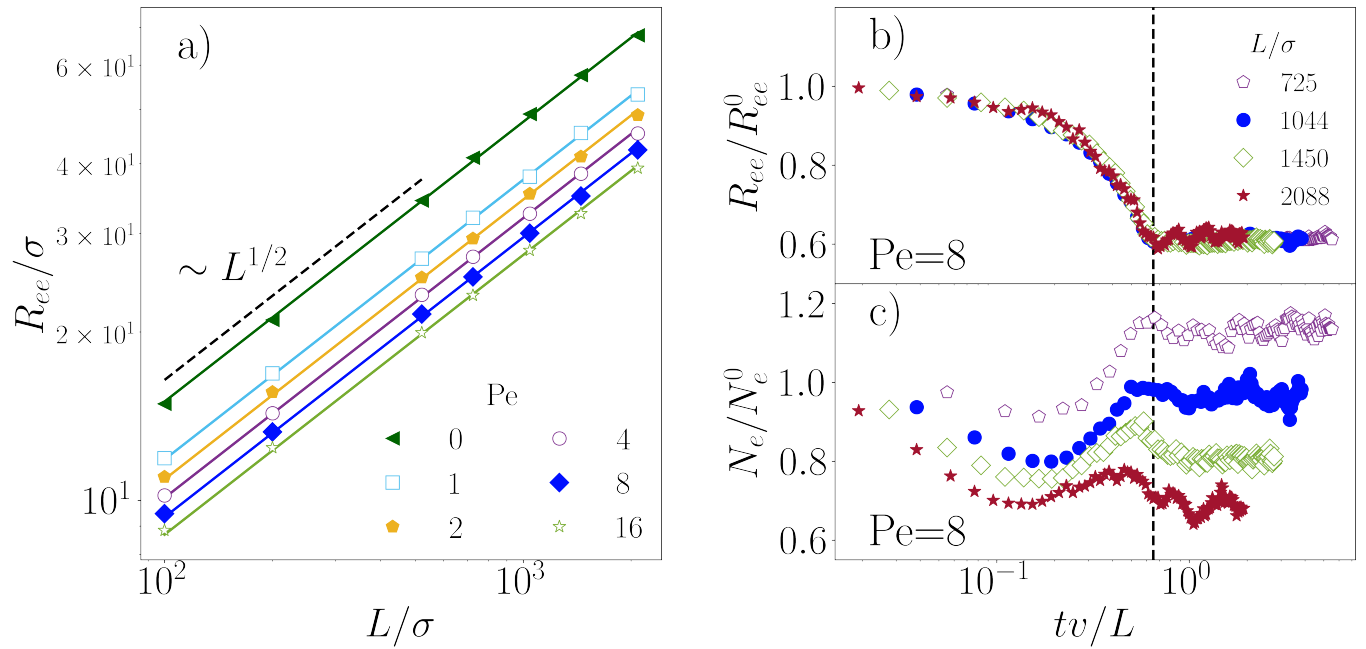

Figure 3. (a) End-to-end distance $R_{ee}$ as a function of polymer size $L/\sigma$ for various Péclet numbers Pe, exhibiting the characteristic end-to-end scaling behavior reminiscent of an ideal polymer chain, $\sim L^{1/2}$. (b) $R_{ee}/R_{ee}^0$ as a function of time for varied $L$ at a fixed Pe = 8, time scaled by $L/v$. (c) Entanglement length $N_e$, normalized by the equilibrium value $N_e^0$ for Pe = 8, as a function of time.

Our investigation of the end-to-end distance $R_{ee}$ of polymer chains at long times reveals a striking consistency: irrespective of the applied Péclet number (Pe), the system exhibits a common scaling law, $R_{ee} \sim L^{1/2}$, similar to ideal polymer solutions [see Fig. 3(a)]. Intriguingly, the prefactor of this scaling relation steadily decreases with increasing Pe, reminiscent of a coil-to-globule transition, observed in dilute active flexible polymer solutions [3]. However, in our complex, densely entangled networks, a true globule-like structure doesn't occur; instead, the $L^{1/2}$ scaling exponent

remains valid across all Pe values, highlighting a consistent entangled behavior in response to activity.

In Fig. 3(b), we explore the temporal evolution of the end-to-end distance for various polymer lengths $L$ at a fixed Péclet number Pe = 8. Given that the $R_{ee} \sim L^{1/2}$ scaling remains valid across for all Péclet numbers, we anticipate the normalized $R_{ee}/R_{ee}^0$ to collapse at long times ($tv/L \gg 1$), as depicted in Fig. 3(b). Furthermore, we note a gradual reduction in $R_{ee}$, persisting until $tv/L \sim 1$. In contrast, we observe a 10% increase in the contour length of the primitive path ($L_{pp}$) at intermediate times ($tv/L \sim 0.1$), as illustrated in Fig. 1(c) of the main text. As a consequence, the entanglement length ($N_e/N_e^0$) is expected to exhibit a 30% decline at $tv/L \sim 0.1$ before ultimately reaching a saturation value at long times [see Fig. 3(c)].

### Viscoelasticity at a fixed polymer length

We explore the complete time-dependent stress autocorrelation functions across a range of Péclet numbers (Pe), while keeping the polymer length fixed at $L/\sigma = 725$. In equilibrium, we find the familiar stress plateau $G_0 = 4\rho k_B T/(5N_e^0)$ [Fig. 4]. However, upon introducing activity, we observe that the short-time behavior of $G(t)$ increases by a factor of 4 compared to its equilibrium counterpart. This emphasizes the active role in shaping the early-time dynamics within the entangled tubes.

Moving on to intermediate times, $t \sim \tau_0$, the grip forces between neighboring polymers intensify, effectively acting as barriers for the individual polymer chains. Consequently, the system struggles to relax, resulting in a remarkable increase in the stress plateau. In fact, for $L/\sigma = 725$, the stress plateau height increases by more than three orders of magnitude [Fig. 4]. This striking phenomenon highlights the pivotal role played by activity-induced grip forces in shaping the viscoelastic responses

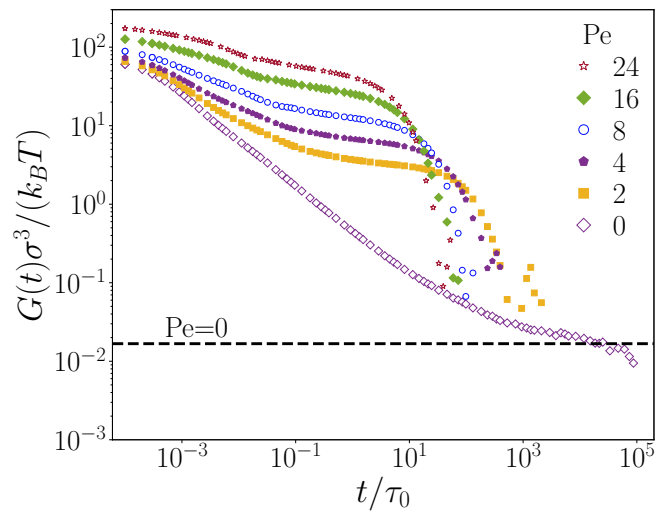

Figure 4. Stress relaxation modulus $G(t)$ as a function of $t/\tau_0$ for $L = 725\sigma$ and varying P'eclet numbers. The dashed line represents the well-established prediction $G_0 = 4\rho k_B T/(5N_e^0)$.

### Viscoelasticity of less entangled systems

To demonstrate the unique nature of the stress plateau enhancement due to activity in entangled solutions, we investigate polymer solutions with shorter polymer lengths, specifically $L = 100\sigma$. In Fig. 5, it becomes evident that the stress plateau is entirely absent from $G(t)$. Instead, the stress relaxation modulus now exhibits a distinct behavior: an initial $\sim t^{-1/2}$ decrease at short times, followed by an eventual exponential decay. This behavior aligns with the predictions of the Rouse model $G(t) \simeq k_B T \rho \, (t/\tau_0)^{-1/2} \, e^{-t/\tau_R}$ ($\tau_R$ is the Rouse time) [4], which describes the relaxation dynamics of polymers in this low-entanglement-regime.

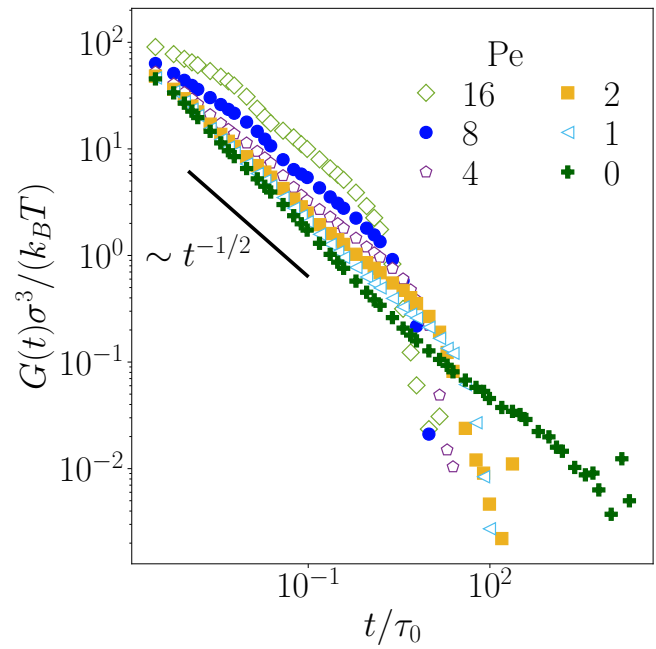

Figure 5. Stress relaxation modulus $G(t)$ as a function of $t/\tau_0$ for $L = 100\sigma$ and varying Péclet numbers. The stress relaxation modulus exhibits an initial decay characterized by $\sim t^{-1/2}$ behavior, followed by a subsequent exponential decay, notably lacking the entangled plateau.

## Movie

The movie (M1.mp4) illustrates the dynamic evolution of primitive paths involving a test polymer (in red) along with its neighboring polymers (in blue) in a simulation setting characterized by Pe = 4, $L = 1450\sigma$, and $\rho^\star = 0.85$. Notably, it reveals an increase in the primitive path, expanding from $L_{pp}/\sigma = 291.3$ to $L_{pp}/\sigma = 348.6$ at intermediate times $tv/L \sim 0.14$. Ultimately, the contour length of the primitive path $L_{pp}$ decreases by 40% compared to its passive counterpart at long times ($tv/L \geq 1$).

* suvendu.mandal@pkm.tu-darmstadt.de
[1] J. D. Dietz and R. S. Hoy, Facile equilibration of well-entangled semiflexible bead–spring polymer melts, J. Chem. Phys. **156**, 10.1063/5.0072386 (2022).
[2] M. Kröger, J. D. Dietz, R. S. Hoy, and C. Luap, The z1+ package: Shortest multiple disconnected path for the analysis of entanglements in macromolecular systems, Comput. Phys. Commun. **283**, 108567 (2023).
[3] V. Bianco, E. Locatelli, and P. Malgaretti, Globulelike conformation and enhanced diffusion of active polymers, Phys. Rev. Lett. **121**, 217802 (2018).
[4] M. Rubinstein and R. H. Colby, Polymer physics, Oxford University Press https://doi.org/10.1002/pi.1472 (2003).


\fi  

\end{document}